\newcommand*{\ATLASLATEXPATH}{latex/}

\ifdefined\InternalMode
\def\ClassOpts{PAPER,UKenglish,texlive=2011,coverpage,block=none}
\else
\def\ClassOpts{cernpreprint,texlive=2011,txfonts,UKenglish}
\fi

\documentclass[\ClassOpts]{\ATLASLATEXPATH atlasdoc}

\usepackage[subfigure=true]{\ATLASLATEXPATH atlaspackage}
\usepackage{\ATLASLATEXPATH atlasbiblatex}

\usepackage{\ATLASLATEXPATH atlascontribute}

\usepackage{\ATLASLATEXPATH atlasphysics}

\graphicspath{{logos/}{figuresRelease/}}

\nolinenumbers

\usepackage{relsize}
\usepackage{Paper_100MeV-defs}
\usepackage{pdfpages}

\AtlasTitle{Charged-particle distributions at low transverse momentum in $\sqrt{s} = 13$~TeV {\itshape pp} interactions measured with the ATLAS detector at the LHC}

\author{The ATLAS Collaboration}

\ifdefined\InternalMode
\AtlasVersion{3.5}
\else
\fi

\AtlasRefCode{STDM-2015-17}

\AtlasJournal{EPJC}
\AtlasJournalRef{Eur. Phys. J. C 76 (2016) 502}
\AtlasDOI{10.1140/epjc/s10052-016-4335-y}

\AtlasAbstract{%
  Measurements of distributions of charged particles produced in proton--proton collisions with a centre-of-mass energy of 13\,TeV are presented. The data were recorded by the ATLAS detector at the LHC and correspond to an integrated luminosity of 151\,$\mu$b$^{-1}$. The particles are required to have a transverse momentum greater than 100\,MeV and an absolute pseudorapidity less than 2.5. The charged-particle multiplicity, its dependence on transverse momentum and pseudorapidity and the dependence of the mean transverse momentum on multiplicity are measured in events containing at least two charged particles satisfying the above kinematic criteria.  The results are corrected for detector effects and compared to the predictions from several Monte Carlo event generators.
}

\ifdefined\InternalMode

\AtlasCoverSupportingNote{Supporting note for the Minimum Bias publication: Charged particle multiplicities in pp interactions in the 100 MeV phase-space measured with the ATLAS detector}{https://cds.cern.ch/record/2102358/}
\AtlasCoverCommentsDeadline{24.04.2016}

\AtlasCoverAnalysisTeam{Alison Lister (Vancouver UBC), Anthony Keith Morley (Sydney), Christoph Eckardt (DESY), Daniela Boerner (Wuppertal), Deepak Kar (Witwatersrand), Douglas Schaefer (CERN), Filip Nechansky (Prague AS), Heather Gray (CERN), Jiri Kvita (Olomouc), Krzysztof Wozniak (Cracow IFJ PAN), Maria Roberta Devesa (Buenos Aires), Oldrich Kepka (Prague AS), Per Edvin Sidebo (Stockholm KTH), Roman Lysak (Prague AS), Thorsten Kuhl (DESY), Valentina Maria Cairo (Cosenza)}

\AtlasCoverEdBoardMember{Claudia Gemme~(chair),  Kirill Prokofiev, Patricia Ward}

\AtlasCoverEgroupEditors{atlas-STDM-2015-17-editors@cern.ch}

\AtlasCoverEgroupEdBoard{atlas-STDM-2015-17-editorial-board@cern.ch}

\else
\fi
\hypersetup{pdftitle={ATLAS draft},pdfauthor={The ATLAS Collaboration}}

\ifdefined\InternalMode

\else
	
\PreprintIdNumber{CERN-EP-2016-099}
\fi
\begin{document}
	
\ifdefined\InternalMode

\else
\nolinenumbers
\fi	

\maketitle



\section{Introduction}
\label{sec:intro}

Measurements of charged-particle distributions in proton–proton ({\itshape pp}) collisions probe the strong interaction in the low-momentum transfer, non-perturbative region of quantum chromodynamics (QCD). In this region, charged-particle interactions are typically described by QCD-inspired models implemented in Monte Carlo (MC) event generators. Measurements are used to constrain the free parameters of these models. An accurate description of low-energy strong interaction processes is essential for simulating single {\itshape pp} interactions and the effects of multiple {\itshape pp} interactions in the same bunch crossing at high instantaneous luminosity in hadron colliders. Charged-particle distributions have been  measured previously in hadronic collisions at various centre-of-mass energies~\cite{STDM-2010-06, Paper8TeV, CMS-QCD-10-004, cmsminbias_2,CMS-QCD-09-010,alice_mb2,Aaltonen:2009ne,Khachatryan:2015jna, Paper500, UA1, UA5}.

\begin{sloppypar}
The measurements presented in this paper use data from {\itshape pp} collisions at a centre-of-mass energy \mbox{$\sqrt{s} = 13 \TeV$} recorded by the ATLAS experiment~\cite{PERF-2007-01} at the Large Hadron Collider (LHC)~\cite{LHC} in 2015, corresponding to an integrated luminosity of 151\,$\mu$b$^{-1}$. The data were recorded during special fills with low beam currents and reduced focusing to give a mean number of interactions per bunch crossing of 0.005. The same dataset and a similar analysis strategy were used to measure distributions of charged particles with transverse momentum $\pt$ greater than 500\,MeV~\cite{Paper500}. This paper extends the measurements to the low-$\pt$ regime of  $\pt > 100$\,MeV.
While this nearly doubles the overall number of particles in the kinematic acceptance, the measurements are rendered more difficult due to multiple scattering and imprecise knowledge of the material in the detector. Measurements in the low-momentum regime provide important information  for the description of the strong interaction in the low-momentum-transfer, non-perturbative region of QCD.
\end{sloppypar}

These measurements use tracks from primary charged particles, corrected for detector effects to the particle level, and are presented as inclusive distributions in a fiducial phase space region. Primary charged 
particles are defined in the same way as in Refs.~\cite{Paper8TeV, Paper500} as charged particles with a mean lifetime $\tau>300$\,ps, either directly 
produced in {\itshape pp} interactions or from subsequent decays of directly produced particles with 
$\tau < 30$\,ps; particles produced from decays of particles with $\tau > 30$\,ps, denoted 
secondary particles, are excluded. Earlier analyses also included charged particles with a mean lifetime of $30 < \tau < 300$\,ps. These are charged strange baryons and have been removed for the present analysis due to their low reconstruction efficiency. For comparison to the earlier measurements, the measured multiplicity at $\eta=0$ is extrapolated to include charged strange baryons. All  primary  charged  particles  are  required  to  have  a  momentum component transverse to the beam direction $\pt>100$\,MeV and absolute pseudorapidity\footnote{ATLAS uses a right-handed coordinate system with its origin at the nominal interaction point (IP) in the centre of the detector
	and the $z$-axis along the beam pipe. The $x$-axis points from the IP to the centre of the LHC ring, and the $y$-axis points upward. Cylindrical coordinates ($r$,$\phi$) are used in the transverse plane, $\phi$ being the azimuthal angle around the beam pipe.
	The pseudorapidity is defined in terms of the polar angle $\theta$ as $\eta= -\ln\tan(\theta/2)$.} $|\eta|<2.5$ to be within the geometrical acceptance of the tracking detector. Each event is required to have at least two primary charged particles. The following observables are measured:


	        $$
	        \dndeta,   \hspace{13pt}
	        \dndpt,   \hspace{13pt}
	        \dNdn   \hspace{13pt}	
	         {\rm and} \hspace{13pt}		
	       \meanptN.
	        $$
	        %
	       
Here \nch\ is the number of primary charged particles within the kinematic acceptance in an event,
	\nev\ is the number of events with $\nch \geq 2$, and \Nch\ is the
	total number of primary charged particles in the kinematic acceptance.
	
The \py~\cite{Sjostrand:2007gs},
\epos~\cite{Porteboeuf:2010um} and \qgsjet~\cite{Qgsjet} MC generators are used to correct the data for detector effects and to compare with particle-level corrected data. 
\py\ and \epos\ both model the effects of colour coherence, which is important in dense parton environments and effectively reduces the number of particles produced in multiple parton-parton interactions.
In \py, the simulation is split into non-diffractive and diffractive processes, the former dominated by $t$-channel gluon exchange and amounting to approximately 80\% of the selected events, and the latter described by a pomeron-based approach~\cite{corke_interleaved_2011}. In contrast, \epos\ implements a parton-based Gribov--Regge~\cite{Drescher:2000ha} theory, an effective field theory describing both hard and soft scattering at the same time. \qgsjet\ is based upon the Reggeon field theory framework~\cite{Reggeon}. The latter two generators do not rely on parton distribution functions (PDFs), as used in \py. Different parameter settings in the models are used in the simulation to reproduce existing experimental data and are referred to as tunes. For \py, the \atwo~\cite{Atwo} tune is based on the \mstwlo\ PDF \cite{Martin:2009iq} while the \monash~\cite{Monash} underlying-event tune uses the \nnpdflo\ PDF \cite{Ball:2012cx} and incorporates updated fragmentation parameters, as well as SPS and Tevatron data to constrain the energy scaling. For \epos, the \lhc~\cite{Pierog:2013ria} tune is used, while for \qgsjet\ the default settings of the generator are applied. Details of the MC generator versions and settings are shown in Table~\ref{tab:MC_samples}. Detector effects are simulated using the GEANT4-based~\cite{Agostinelli:2002hh} ATLAS simulation framework~\cite{simulation}.

\begin{table}[htb]
	\caption{ Summary of MC generators used to compare to the corrected data. The generator, its version, the corresponding tune and the parton distribution function are given.}
	\begin{center}
		\begin{tabular}{l|lll}
			Generator & Version & Tune & PDF \\ \midrule
			\py & 8.185 & \atwo & \mstwlo \\
			\py & 8.186 & \monash &\nnpdflo \\
			\epos & LHCv3400 & \lhc & -- \\
			\qgsjet & \textsc{II-04} & default & --
		\end{tabular}
	\end{center}
	\label{tab:MC_samples}
\end{table} 
\section{ATLAS detector}
\label{sec:detector}

The ATLAS detector covers nearly the whole solid angle around the collision point and includes tracking detectors, calorimeters and muon chambers. This measurement uses information from the inner detector and the trigger system, relying on the minimum-bias trigger scintillators (MBTS).

The inner detector covers the full range in $\phi$ and $|\eta| < 2.5$. It consists of the silicon pixel detector (pixel), the silicon microstrip detector (SCT) and the transition radiation straw-tube tracker (TRT). These are located around the interaction point spanning radial distances of 33--150~mm, 299--560~mm and 563--1066~mm respectively. The barrel (each end-cap) consists of four (three) pixel layers, four (nine) double-layers of silicon microstrips and 73 (160) layers of TRT straws.  During the LHC long shutdown 2013--2014, a new innermost pixel layer, the insertable B-layer (IBL)~\cite{Capeans:1291633,ATLAS:1451888}, was installed around a new smaller beam-pipe. The smaller radius of 33~mm and the reduced pixel size of the IBL result in improvements of both the transverse and longitudinal impact parameter resolutions. Requirements on an innermost pixel-layer hit and on impact parameters strongly suppress the number of tracks from secondary particles. A track from a charged particle passing through the barrel typically has  12 measurement points (hits) in the pixel and SCT detectors. The inner detector is located within a solenoid that provides an axial 2\,T magnetic field.

A two-stage trigger system is used: a hardware-based level-1 trigger (L1) and a software-based high-level trigger (HLT). The L1 decision provided by the MBTS detector is used for this measurement. The scintillators are installed on either side of the interaction point in front of the liquid-argon end-cap calorimeter cryostats at $z=\pm3.56$~m and segmented into two rings in pseudorapidity ($2.07<|\eta|<2.76$ and $2.76<|\eta|<3.86$). The inner (outer) ring consists of eight (four) azimuthal sectors, giving a total of 12 sectors on each side. The trigger used in this measurement requires at least one signal in a scintillator on one side  to be above threshold.

\section{Analysis}

The analysis closely follows the strategy described in Ref.~\cite{Paper500}, but modifications for the low-\pt region are applied where relevant. 

\subsection{Event and track selection}
Events are selected from colliding proton bunches using the MBTS trigger described above. Each event is required to contain a primary vertex~\cite{Piacquadio:2008zzb}, reconstructed from at least two tracks with a minimum \pt\ of 100\,MeV. To reduce contamination from events with more than one interaction in a bunch crossing, events with a second vertex containing four or more tracks are removed. The contributions from non-collision background events and the fraction of events where two interactions are reconstructed as a single vertex have been studied in data and are found to be negligible.

\begin{figure*} [htb!]
	\begin{center}
			\includegraphics[width=0.55\textwidth]{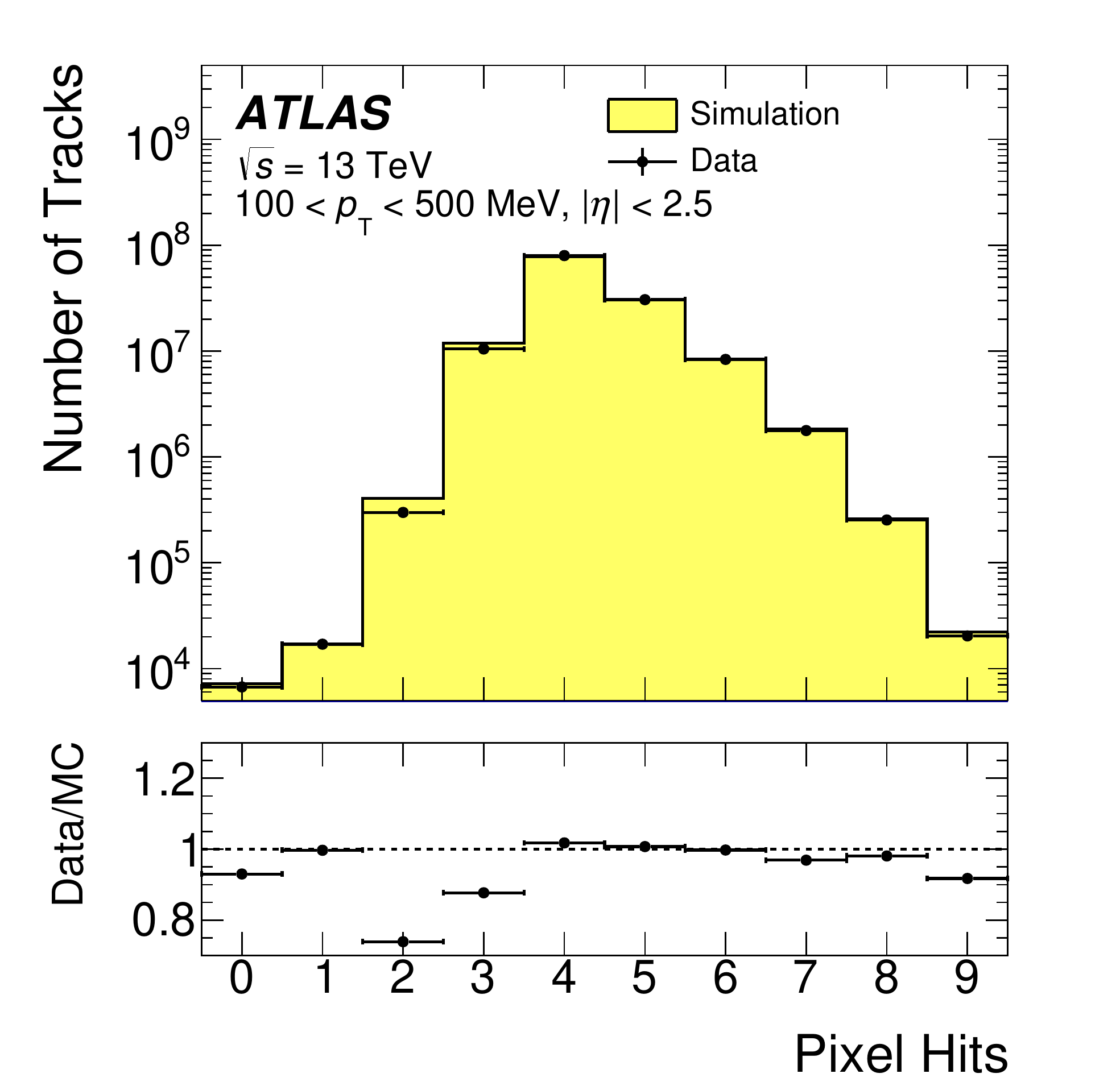}

	\end{center}
	\caption{ Comparison between data and \py~\atwo\ simulation for the distribution of the number of pixel hits associated with a track. The distribution is shown before the requirement on the number of pixel hits is applied, for tracks with $100 < \pt < 500\MeV$ and $|\eta| < 2.5$. The error bars on the points are the statistical uncertainties of the data. The lower panel shows the ratio of data to MC prediction.
		\label{fig:Perf_pixel}}
\end{figure*}

Track candidates are reconstructed in the pixel and SCT detectors and extended to include measurements in the TRT~\cite{newt1,newt2}. A special configuration of the track reconstruction algorithms was used for this analysis to reconstruct low-momentum tracks with good efficiency and purity. The purity is defined as the fraction of  selected tracks that are also primary tracks  with a transverse momentum of at least 100\,MeV and an absolute pseudorapidity less than 2.5. The most critical change with respect to the 500\,MeV analysis~\cite{Paper500}, besides lowering the $\pta$ threshold to 100\,MeV, is reducing the requirement on the minimum number of silicon hits from seven to five. All tracks, irrespective of their transverse momentum, are reconstructed in a single pass of the track reconstruction algorithm. Details of the performance of the track reconstruction in the 13\,TeV data and its simulation can be found in Ref.~\cite{ATL-PHYS-PUB-2015-018}. Figure~\ref{fig:Perf_pixel} shows the comparison between data and simulation in the distribution of the number of pixel hits associated with a track for the low-momentum region. Data and simulation agree reasonably well given the known imperfections in the simulation of inactive pixel modules. These differences are taken into account in the systematic uncertainty on the tracking efficiency by comparing the efficiency of the pixel hit requirements in data and simulation after applying all other track selection requirements.

 Events are required to contain at least two selected tracks satisfying the following criteria: $\pta>100\MeV$ and  $|\eta | < 2.5$; at least one pixel hit and an innermost pixel-layer hit if expected;\footnote{A hit is expected if the extrapolated track crosses an known active region of a pixel module. If an innermost pixel-layer hit is not expected, a next-to-innermost pixel-layer hit is required if expected.} at least two, four or six SCT hits for $\pt < 300\MeV$, $< 400\MeV$ or $> 400\MeV$ respectively, in order to account for the dependence of track length on $\pt$; $| \dzerobs |< 1.5$\,mm,  where  the transverse impact parameter \dzerobs\ is calculated with respect to the measured beam line (BL); and $|\zzsint| < 1.5$\,mm, where $\zBL$ is the difference between the longitudinal position of the track along the beam line at the point where $\dzerobs$ is measured and the longitudinal position of the primary vertex and $\theta$ is the polar angle of the track. High-momentum tracks with mismeasured \pta\ are removed by requiring the track-fit $\chi^2$ probability to be larger than 0.01 for tracks with $\pta>10\GeV$. In total $9.3\times10^{6}$ events pass the selection, containing a total of $3.2\times10^{8}$ selected tracks.

\subsection{Background estimation}
\label{sec:bkg_estimate}
\begin{figure*} [htb!]
	\begin{center}
			\includegraphics[width=0.55\textwidth]{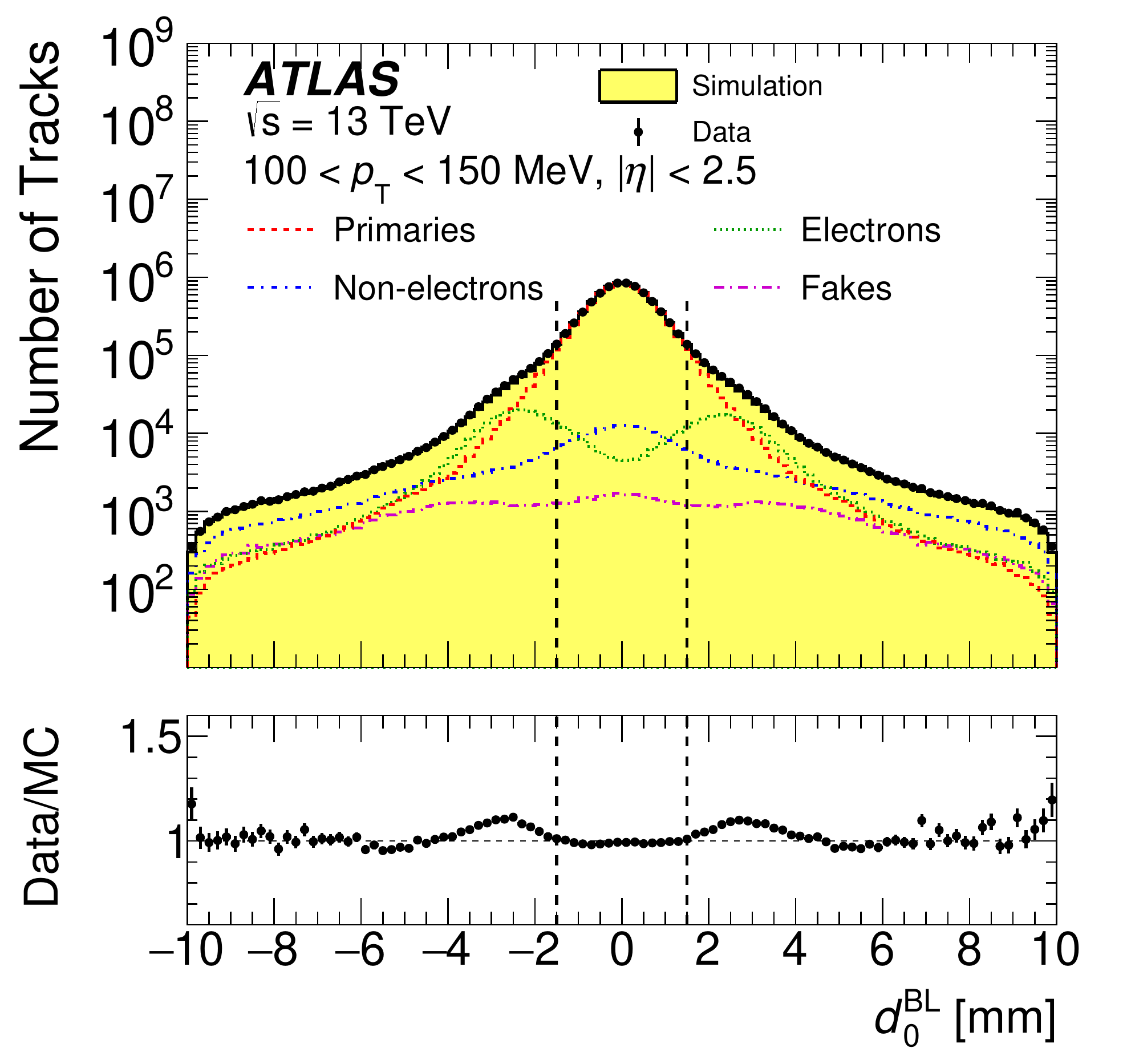}
	\end{center}
	\caption{ Comparison between data and \py~\atwo\ simulation for the transverse impact parameter $\dzerobs$ distribution. The $\dzerobs$ distribution is shown for $100 < \pt < 150\MeV$ without applying the cut on the transverse impact parameter. The position where the cut is applied is shown as dashed black lines at $\pm1.5$\,mm. The simulated $\dzerobs$ distribution is normalised to the number of tracks in data and the separate contributions from primary, fake, electron and non-electron tracks are shown as lines using various combinations of dots and dashes. The secondary particles are scaled by the fitted fractions as described in the text. The error bars on the points are the statistical uncertainties of the data. The lower panel shows the ratio of data to MC prediction.
	\label{fig:Perf_d0}}
\end{figure*}

Background contributions to the tracks from primary particles include fake tracks (those formed by a random combination of hits), strange baryons and secondary particles. These contributions are subtracted on a statistical basis from the number of reconstructed tracks before correcting for other detector effects. The contribution of fake tracks, estimated from simulation, is at most 1\% for all \pt and \eta intervals with a relative uncertainty of $\pm$50\% determined from dedicated comparisons of data with simulation~\cite{fakeUnc}. Charged strange baryons with a mean lifetime $30 < \tau < 300$\,ps are treated as background, because these particles and their decay products have a very low reconstruction efficiency. Their contribution is estimated from \epos, where the best description of this strange baryon contribution is expected~\cite{Paper500}, to be below 0.01\% on average, with the fraction increasing with track \pt to be $(3\pm1)\%$ above 20\,GeV. The fraction is much smaller at low \pt due to the extremely low track reconstruction efficiency.
The contribution from secondary particles is estimated by performing a template fit to the distribution of the track transverse impact parameter \dzerobs, using templates for primary and secondary particles created from \py\ \atwo\ simulation.
All selection requirements are applied except that on the transverse impact parameter. The shape of the transverse impact parameter distribution differs for electron and non-electron secondary particles, as the $\dzerobs$ reflects the radial location at which the secondaries were produced. The processes for conversions and hadronic interactions are rather different, which leads to differences in the radial distributions. The electrons are more often produced from conversions in the beam pipe. Furthermore, the fraction of electrons increases as \pt decreases. Therefore, separate templates are used for electrons and non-electron secondary particles in the region $\pT < 500$\,MeV. The rate of secondary tracks is the sum of these two contributions and is measured with the fit. The background normalisation for fake tracks and strange baryons is determined from the prediction of the simulation.
 The fit is performed in nine \pt\ intervals, each of width 50\,MeV, in the region $4<|\dzerobs|<9.5$ mm. The fitted distribution for $100 < \pt < 150\MeV$ is shown in Figure~\ref{fig:Perf_d0}. For this $\pt$ interval, the fraction of secondary tracks within the region $|\dzerobs| < 1.5$\,mm is measured to be $(3.6\pm0.7)\%$, equally distributed between electrons and non-electrons. For tracks with $\pt > 500\MeV$, the fraction of secondary particles is measured to be $(2.3\pm0.6)\%$; these are mostly non-electron secondary particles.  The uncertainties are evaluated by using different generators to estimate the interpolation from the fit region to $|\dzerobs|<1.5$ mm, changing the fit range  and checking the $\eta$ dependence of the fraction of tracks originating from secondaries. This last study is performed by fits integrated over different \eta ranges, because the $\eta$ dependence could be different in data and simulation, as most of the secondary particles are produced in the material of the detector. The systematic uncertainties arising from imperfect knowledge of the passive material in the detector are also included; these are estimated using the same material variations as used in the estimation of the uncertainty on the tracking efficiency, described in Section~\ref{sec:trk_reco_eff}.

\subsection{Trigger and vertex reconstruction efficiency}
The trigger efficiency $\varepsilon_{\mathrm{trig}}$  is measured in a data sample recorded using a control trigger which selected events randomly at L1 only requiring that  the beams are colliding in the ATLAS detector. The events are  then filtered at the HLT by requiring at least one reconstructed track with $\pT > 200\MeV$. 
The efficiency $\varepsilon_{\mathrm{trig}}$ is defined as the ratio of events that are accepted by both the control and the MBTS trigger to all events accepted by the control trigger. It is measured as a function of the number of selected tracks with the requirement on the longitudinal impact parameter removed, $n_{\textrm{sel}}^{\textrm{no-z}}$. The trigger efficiency increases from $96.5^{+0.4}_{-0.7}$\,\% for events with $n_{\textrm{sel}}^{\textrm{no-z}}=2$
, to $(99.3\pm0.2)\%$ for events with $n_{\textrm{sel}}^{\textrm{no-z}}\geq4$. The quoted uncertainties include statistical and systematic uncertainties. The systematic uncertainties are estimated from the difference between the trigger efficiencies measured on the two sides of the detector, and the impact of beam-induced background; the latter is estimated using events recorded when only one beam was present at the interaction point, as described in Ref.~\cite{Paper500}.

The vertex reconstruction efficiency $\varepsilon_{\textrm{vtx}}$ is determined from data by calculating the ratio of the number of triggered events with a reconstructed vertex to the total number of all triggered events. The efficiency, measured as a function of  $n_{\textrm{sel}}^{\textrm{no-z}}$, is approximately 87\% for events with $n_{\textrm{sel}}^{\textrm{no-z}}=2$  and rapidly rises to 100\% for events with $n_{\textrm{sel}}^{\textrm{no-z}} > 4$. For events with $n_{\textrm{sel}}^{\textrm{no-z}}=2$, the efficiency is also parameterised as a function of the difference between the longitudinal impact parameter of the two tracks ($\Delta z_{\textrm{tracks}}$). This efficiency decreases roughly linearly from 91\% at $\Delta z_{\textrm{tracks}} = 0$\,mm to 32\% at $\Delta z_{\textrm{tracks}} = 10$\,mm. The systematic uncertainty is estimated from the difference between the vertex reconstruction efficiency measured before and after beam-background removal and found to be negligible.

\subsection{Track reconstruction efficiency}
\label{sec:trk_reco_eff}	
The primary-track reconstruction efficiency \trkeff\ is determined from simulation. The efficiency is parameterised in two-dimensional bins of \pt and $\eta$, and
is defined as:

$$
\trkeff (p_\mathrm{T},\eta) \ = \ \frac{N^\mathrm{matched}_\mathrm{rec}(p_\mathrm{T},\eta)}
                                   {N_\mathrm{gen}(p_\mathrm{T},\eta)},
$$

where  \pta\ and $\eta$ are generated particle properties,
$N^\mathrm{matched}_\mathrm{rec}(p_\mathrm{T},\eta)$ is the number of
reconstructed tracks matched to generated primary charged particles 
and $N_\mathrm{gen}(p_\mathrm{T},\eta)$ is the number of generated
primary charged particles in that kinematic region.
A track is matched to a generated particle if the weighted fraction of track hits originating from that particle exceeds 50\%. The hits are weighted such that hits in all subdetectors have the same weight in the sum, based on the number of expected hits and the resolution of the individual subdetector.  For $100 < \pt <  125\MeV$ and integrated over $\eta$, the primary-track reconstruction efficiency is 27.5\%. In the analysis using tracks with $\pt > 500\MeV$~\cite{Paper500}, a data-driven correction to the efficiency was evaluated in order to account for material effects in the $|\eta|>1.5$ region. This correction to the efficiency is not applied in this analysis due to the large uncertainties of this method for low-momentum tracks, which are larger than the uncertainties in the material description. 

\begin{figure*} [htb!]
	\begin{center}
		\subfigure[] {
			\includegraphics[width=0.48\textwidth]{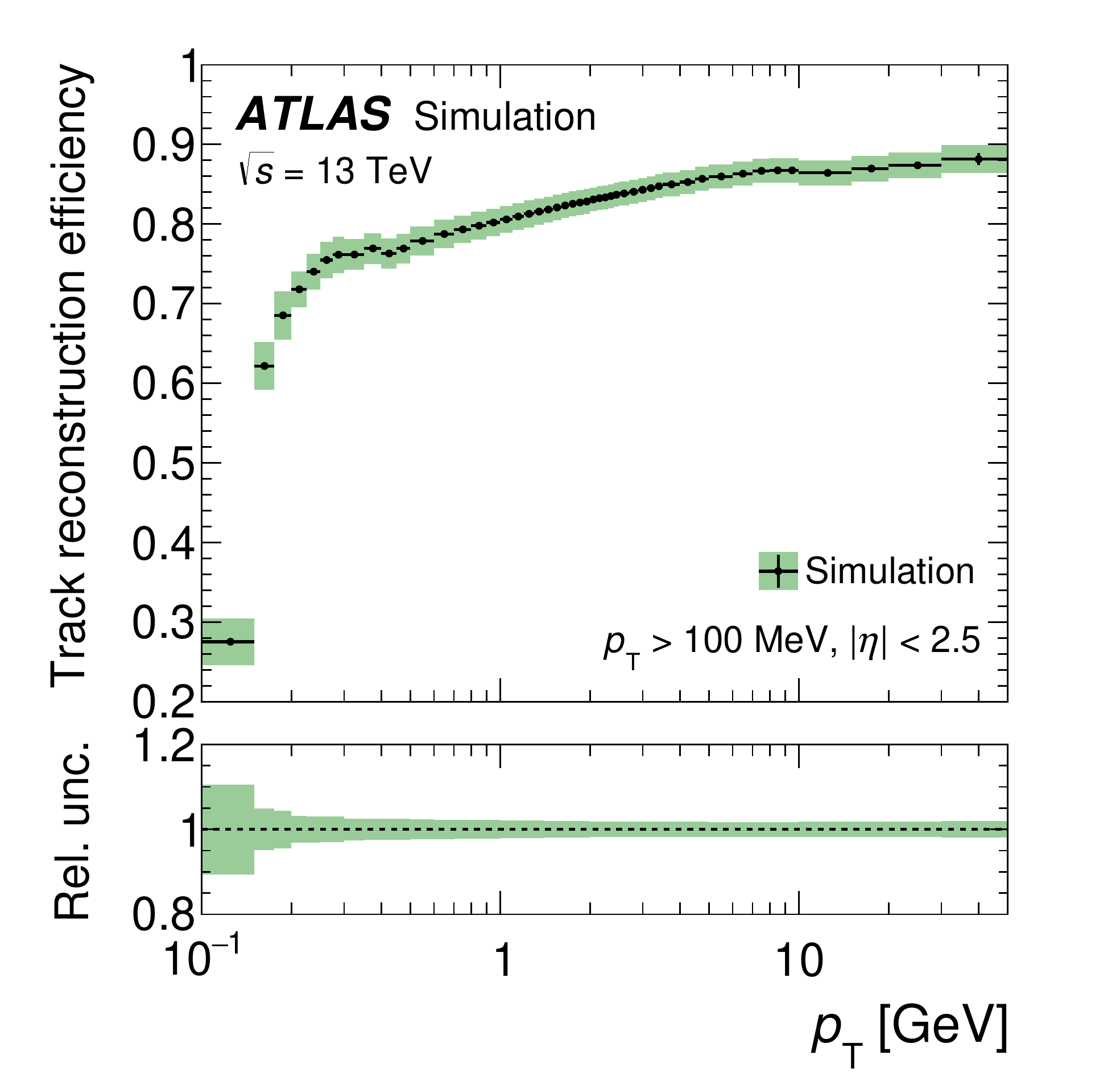}
		}
		\subfigure[] {
			\includegraphics[width=0.48\textwidth]{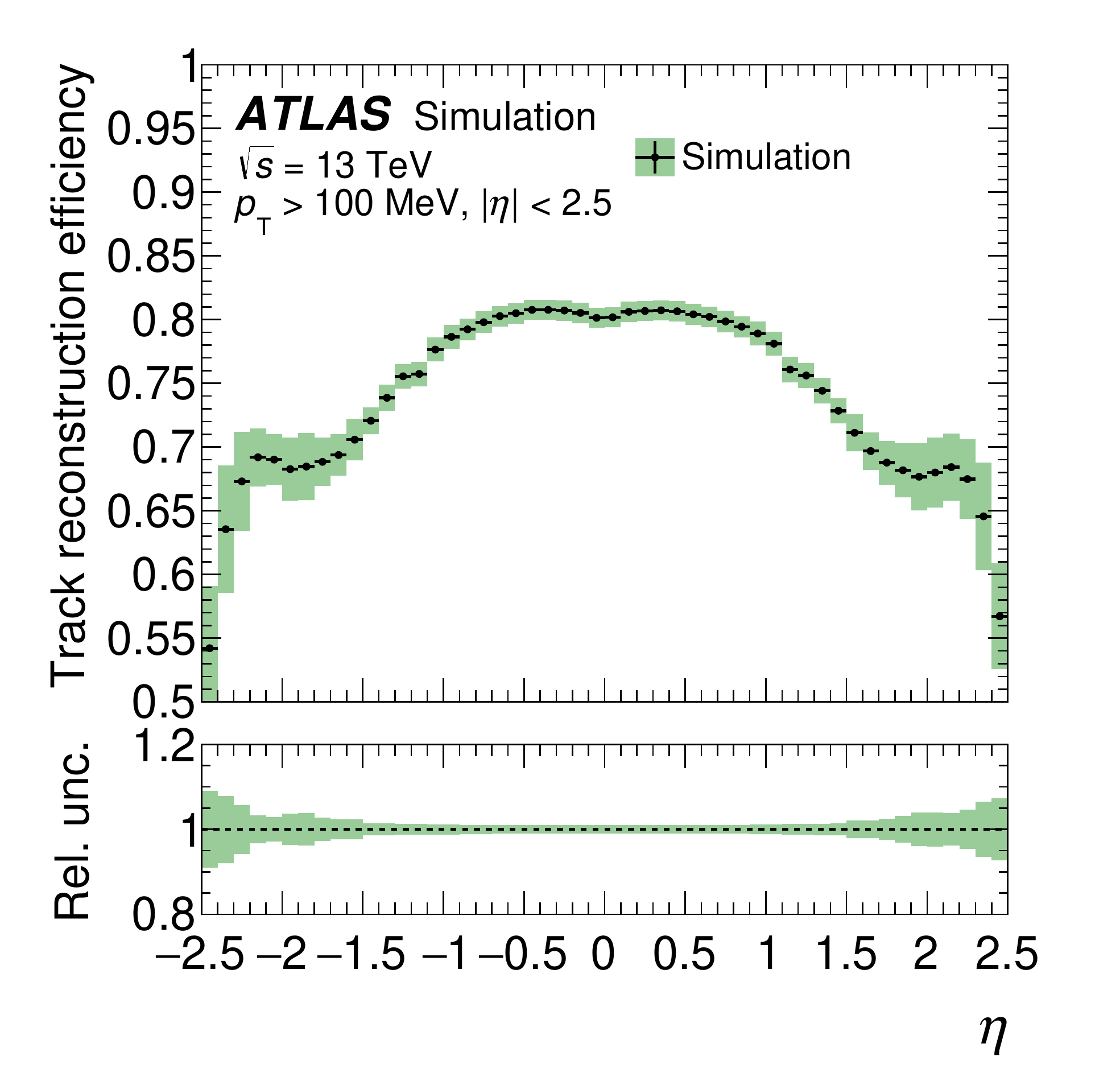}
		}
	\end{center}
	\caption{Track reconstruction efficiency as a function of (a) transverse momentum \pt and of (b) pseudorapidity \eta  for selected tracks with \pt>100\,MeV and $|\eta|<2.5$ as predicted by \py~\atwo\ and single-particle simulation. The statistical uncertainties are shown as vertical bars, the sum in quadrature of statistical and systematic uncertainties as shaded areas.}
	\label{fig:trk_eff_pt}
\end{figure*}

The dominant uncertainty in the track reconstruction efficiency arises from imprecise knowledge of the passive material in the detector. This is estimated by evaluating the track reconstruction efficiency in dedicated simulation samples with increased detector material. The total uncertainty in the track reconstruction efficiency due to the amount of material is calculated as the linear sum of the contributions of 5\%
additional material in the entire inner detector, 10\% additional
material in the IBL and 50\% additional material in the pixel services region at
$|\eta|>1.5$. The sizes of the variations are estimated from studies of the rate of photon conversions, of hadronic interactions, and of tracks lost due to interactions in the pixel services~\cite{sysVarTrk}. The resulting uncertainty in the track reconstruction efficiency is 1\% at low $|\eta|$ and high $\pt$ and up to 10\%  for higher $|\eta|$ or for lower $\pt$. The systematic uncertainty arising from the track selection requirements is studied by comparing the efficiency of each requirement in data and simulation. This results in an uncertainty of 0.5\% for all $\pt$ and $\eta$. The total uncertainty in the track reconstruction efficiency is obtained by adding all effects in quadrature. The track reconstruction efficiency is shown as function of \pt and \eta in Figure~\ref{fig:trk_eff_pt}, including all systematic uncertainties. The efficiency is calculated using the \py~\atwo\ and single-particle simulation. Effectively identical results are obtained when using the prediction from \epos\ or \py~\monash.

\subsection{Correction procedure and systematic uncertainties}
The data are corrected to obtain inclusive spectra for primary charged particles satisfying the particle-level phase space requirement. The inefficiencies due to the trigger selection and vertex reconstruction are applied to all distributions as event weights:
	
\begin{equation}
	        w_\textrm{ev}(\nsel^{\textrm{no-z}}, \Delta z_\textrm{tracks}) = \frac{1}{\varepsilon_\textrm{trig}(\nsel^{\textrm{no-z}})}\cdot\frac{1}{\varepsilon_\textrm{vtx}(\nsel^{\textrm{no-z}}, \Delta z_\textrm{tracks})}.
	\end{equation}

Distributions of the selected tracks are corrected for inefficiencies in the track reconstruction with a track weight using the tracking efficiency ($\varepsilon_\textrm{trk}$) and after subtracting the fractions of fake tracks ($f_\textrm{fake}$), of strange baryons ($f_\textrm{sb}$), of secondary particles ($f_\textrm{sec}$) and of particles outside the kinematic range ($f_\textrm{okr}$):
	 
\begin{equation}
w_\textrm{trk}(\pt, \eta) = \frac{1}{\varepsilon_\textrm{trk}(\pt, \eta)} \cdot
        \left[ 1 - f_\textrm{fake}(\pt, \eta) - f_\textrm{sb}(\pt, \eta) - f_\textrm{sec}(\pt, \eta) - f_\textrm{okr}(\pt, \eta)   \right].
        \label{eq:weight}
	\end{equation}

These distributions are estimated as described in Section~\ref{sec:bkg_estimate} except that the fraction of particles outside the kinematic range whose reconstructed tracks enter the kinematic range is estimated from simulation. This fraction is largest at low $\pt$ and high $|\eta|$. At $\pT = 100$\,MeV and $|\eta| = 2.5$, 11\% of the particles enter the kinematic range and are subtracted as described in Formula~\ref{eq:weight} with a relative uncertainty  of $\pm$ 4.5\%.

The  \pt\ and \eta\ distributions are corrected by the event and track weights, as discussed above. In order to correct for resolution effects, an iterative Bayesian unfolding~\cite{D'Agostini:1994zf} is additionally applied to the \pt\ distribution. The response matrix used to unfold the data is calculated from \py~\atwo\ simulation, and six iterations are used; this is the smallest number of iterations after which the process is stable. The statistical uncertainty is obtained using pseudo-experiments. For the \eta\ distribution, the resolution is smaller than the bin width and an unfolding is therefore unnecessary.
After applying the event weight, the Bayesian unfolding is applied to the multiplicity distribution in order to correct from the observed track multiplicity to the multiplicity of primary charged particles, and therefore the track reconstruction efficiency weight does not need to be applied.
The total number of events, $N_\textrm{ev}$, is defined as the integral of the multiplicity distribution after all corrections are applied and is used to normalise  the distributions. The dependence of \meanpt\ on \nch\ is obtained by first separately correcting the total number of tracks and $\sum_{i}\pt(i)$ (the scalar sum of the track \pt of all tracks with \pt\ > 100\,MeV in one event), both versus the number of primary charged particles. After applying the correction to all events using the event and track weights, both distributions are unfolded separately. The ratio of the two unfolded distributions gives the dependence of \meanpt\ on \nch.


\begin{table}[htb]
	\caption{Summary of the systematic uncertainties in the \eta, \pt, \nch\ and \meanpt\ vs. \nch\ observables. The uncertainties are given at the minimum and the maximum of the phase space.}

	\centering
	\vspace{5pt}
	\setlength{\tabcolsep}{10pt}
	\begin{tabular}{r|cccc}
		        Distribution &    $\dndetaabs$    &     \dndpt      &           \dNdn           &          \meanptN           \\ [5pt]
		               Range & 0--2.5 & 0.1--50\,\GeV &         2--250          &          0--160\,\GeV           \\ [3pt] \midrule
		Track reconstruction &   1\%--7\%    &   1\%--6\%    & 0\%--$^{+38\%}_{-20\%}$ &        0\%--0.7\%         \\[3pt]
		   Track background &     0.5\%     &  0.5\%--1\%   &  0\%--$^{+7\%}_{-1\%}$  &        0\%--0.1\%         \\[3pt]
		        \pt spectrum &      --       &       --        &  0\%--$^{+3\%}_{-9\%}$  & 0\%--$^{+0.3\%}_{-0.1\%}$ \\[3pt]
		         Non-closure & 0.4\%--1\%  &   1\%--3\%    &        0\%--4\%         &        0.5\%--2\%         \\[3pt]
	\end{tabular}
	\label{tab:sys}
\end{table}

A summary of the systematic uncertainties is given in Table~\ref{tab:sys}\ for all observables. The dominant uncertainty is due to material effects on the track reconstruction efficiency. 
Uncertainties due to imperfect detector alignment are taken into account and are less than 5\% at the highest track \pt values. In addition, resolution effects on the transverse momentum can result in low-$\pt$ particles being reconstructed as high-\pT tracks. All these effects are considered as systematic uncertainty on the track reconstruction. The track background uncertainty is dominated by systematic effects in the estimation of the contribution from secondary particles. The track reconstruction efficiency determined in simulation can differ from the one in data if the \pt spectrum is different for data and simulation, as the efficiency depends strongly on the track \pt. This effect can alter the number of primary charged particles and is taken into account as a systematic uncertainty on the multiplicity distribution and \meanpt\ vs \nch.
The non-closure systematic uncertainty is estimated from differences in the unfolding results using \py\ \atwo\ and \epos\ simulations. For this, all combinations of these MC generators are used to simulate the distribution and the input to the unfolding.

\vspace{-1em}
\section{Results}
\label{sec:result}

\vspace{-0.5em}

\begin{figure*} [h!]
	\begin{center} 
		\subfigure[]{
			\includegraphics[width=0.43\textwidth]{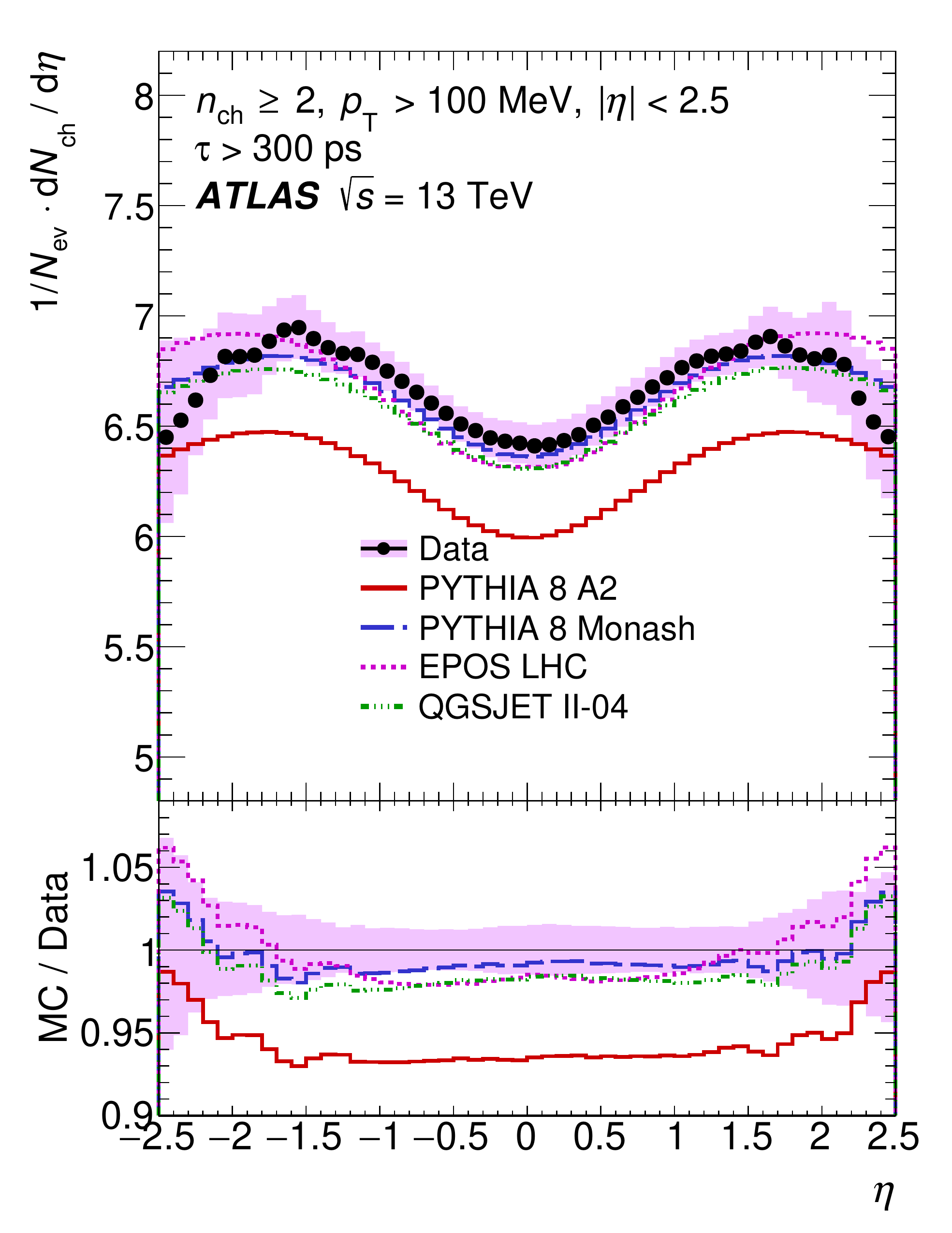}
			\label{fig:dneta}
		}
		\subfigure[]{
			\includegraphics[width=0.43\textwidth]{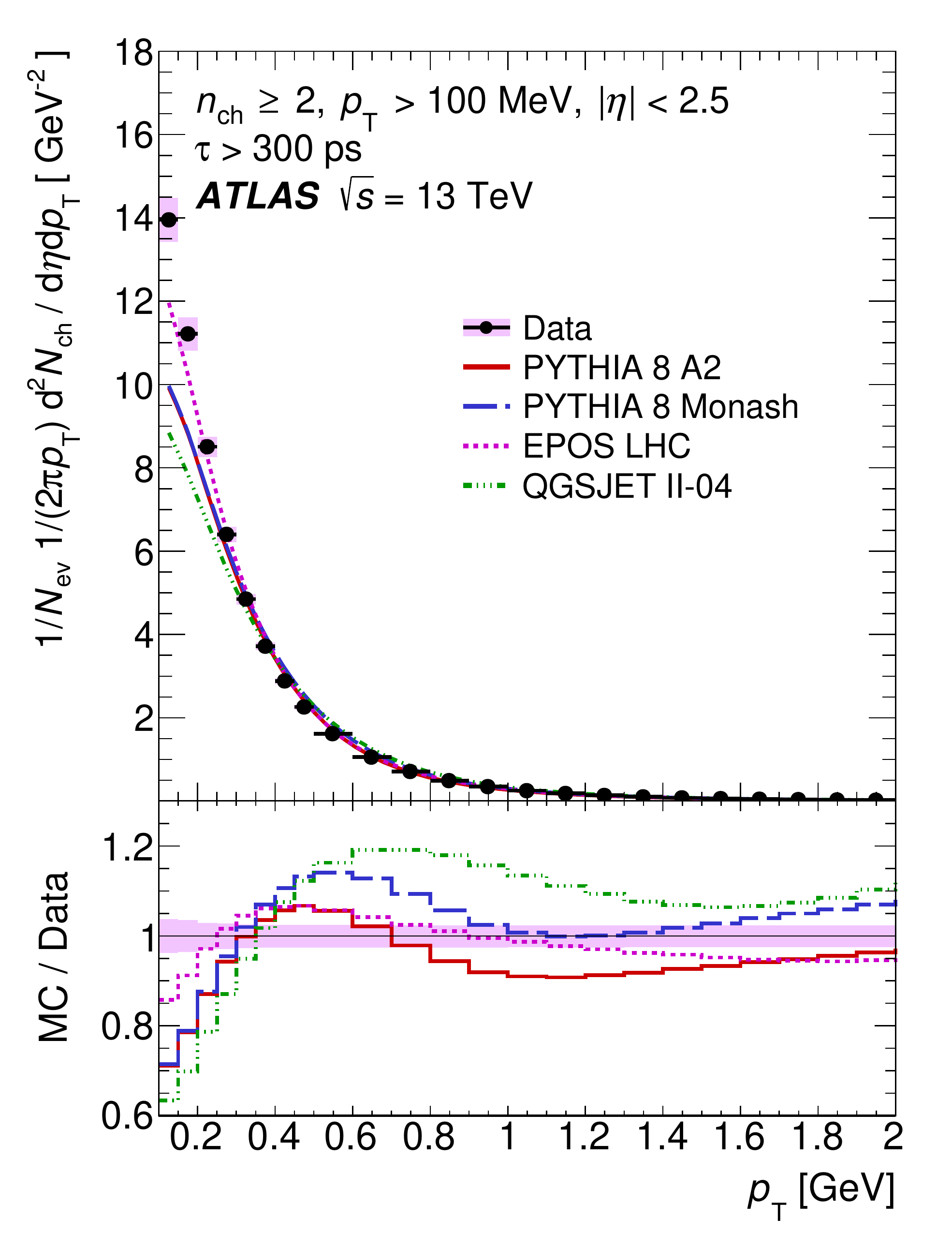}
			\label{fig:dndp}
		}
		\subfigure[]{
			\includegraphics[width=0.43\textwidth]{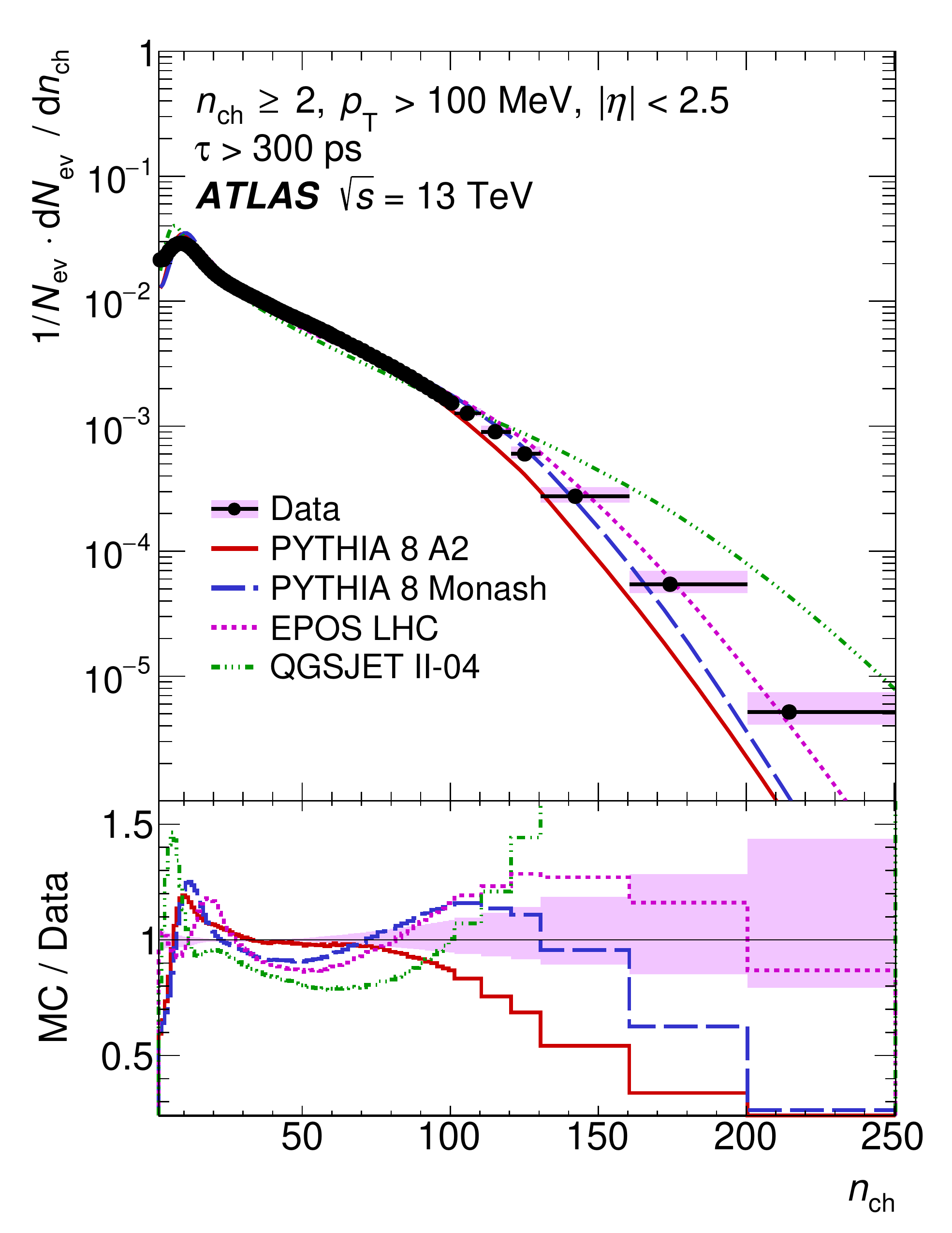}
			\label{fig:dNdnch}
		}
		\subfigure[]{
			\includegraphics[width=0.43\textwidth]{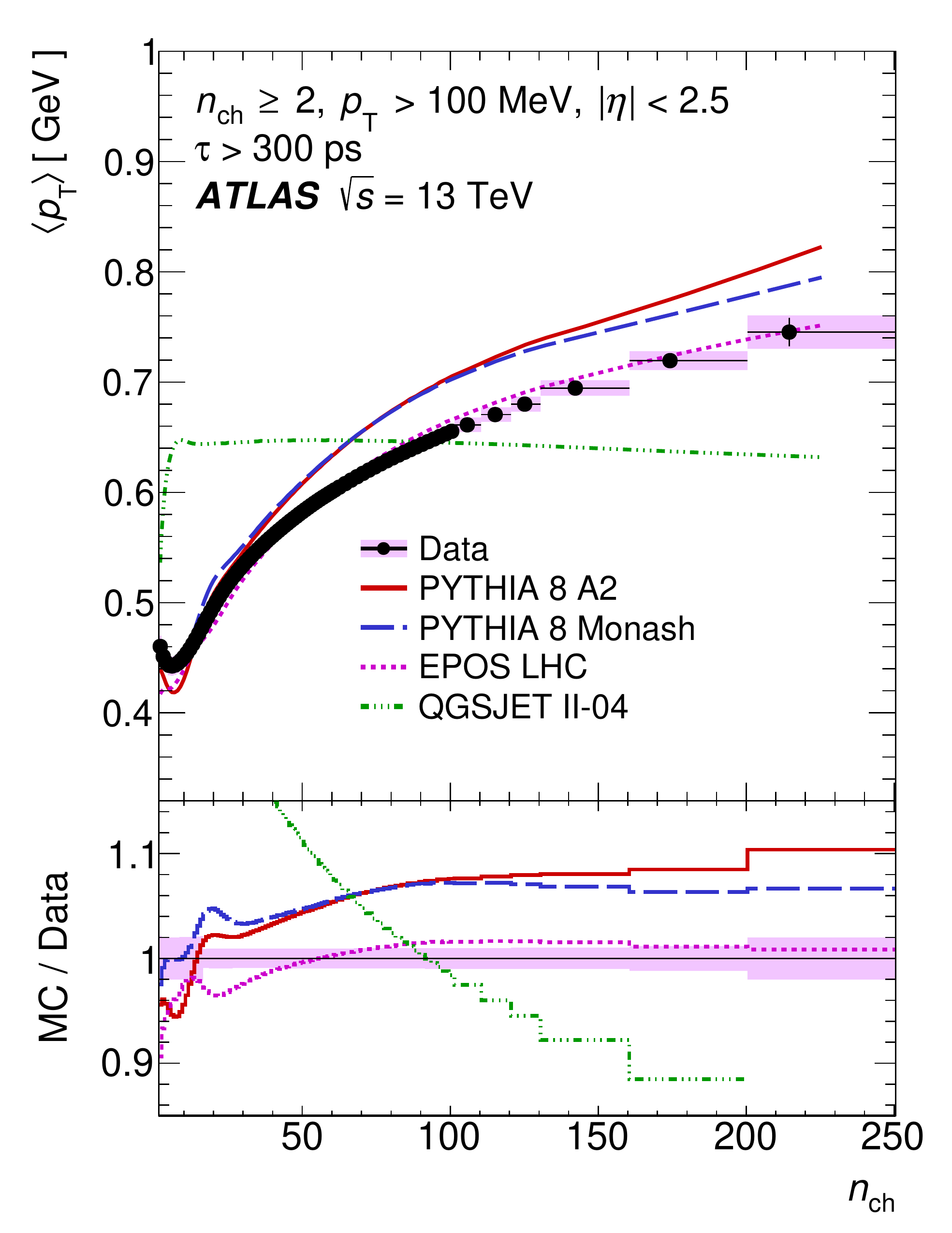}
			\label{fig:mnpt}
		}
	\end{center}
	\caption{Primary charged-particle multiplicities as a function of (a) pseudorapidity $\eta$ and (b) transverse momentum  \pt, (c) the primary charged-particle multiplicity \nch\ and (d) the mean transverse momentum \meanpt\ versus \nch\ for events with at least two primary charged particles with $\pt>100\MeV$ and $|\eta|<2.5$, each with a lifetime $\tau > 300$\,ps.
		 The black dots represent the data and the coloured curves the different MC model predictions.The vertical bars represent the statistical uncertainties, while the shaded areas show statistical and systematic uncertainties added in quadrature. The lower panel in each figure shows the ratio of the MC simulation to data. As the bin centroid is different for data and simulation, the values of the ratio correspond to the averages of the bin content.
		}
	\label{fig:results}
\end{figure*}

The measured charged-particle multiplicities in events containing at least two charged particles with $\mbox{\pt > 100\,\MeV}$ and $|\eta|$ < 2.5 are shown in Figure~\ref{fig:results}. The corrected data are compared to predictions from various generators. In general, the systematic uncertainties are larger than the statistical uncertainties.

Figure~\ref{fig:dneta} shows the charged-particle multiplicity as a function of the pseudorapidity \eta. \py\ \monash, \epos\ and \qgsjet\ give a good description for $|\eta|<1.5$.  The prediction from \py~\atwo\ has the same shape as predictions from the other generators, but lies below the data.

The charged-particle transverse momentum is shown in Figure~\ref{fig:dndp}. \epos\ describes the data well for $\pt > 300\MeV$. For $\pt < 300\MeV$, the data are underestimated by up to 15\%. The other generators show similar mismodelling at low momentum but with larger discrepancies up to 35\% for \qgsjet. In addition, they mostly overestimate the charged-particle multiplicity for $\pt > 400\MeV$; \py~\atwo\ overestimates only in the intermediate \pt region and underestimates the data slightly for $\pt > 800\MeV$.

Figure~\ref{fig:dNdnch} shows the charged-particle multiplicity. Overall, the form of the measured distribution is reproduced reasonably by all models. \py~\atwo\ describes the data well for $30<\nch<80$, but underestimates it for higher \nch.  For $30<\nch<80$, \py~\monash, \epos\ and \qgsjet\ underestimate the data by up to 20\%. \py~\monash\ and \epos\ overestimate the data for $\nch>80$ and drop below the measurement in the high-\nch\ region, starting from $\nch>130$ and $\nch>200$ respectively. \qgsjet\ overestimates the data significantly for $\nch>100$.

The mean transverse momentum versus the primary charged-particle multiplicity is shown in Figure~\ref{fig:mnpt}. It increases towards higher \nch, as modelled by a colour reconnection mechanism in \py\ and by the hydrodynamical evolution model in \epos. The \qgsjet\ generator, which has no model for colour coherence effects, describes the data poorly. For low \nch, \py\ \atwo\ and \epos\ underestimate the data, where \py\ \monash\ agrees within the uncertainties. For higher \nch\ all generators overestimate the data, but for $\nch > 40$, there is a constant offset for  both \py\ tunes, which describe the data to within 10\%. \epos\ describes the data reasonably well and to within 2\%.

\begin{figure*}[h!]
	\begin{center}
		\includegraphics[width=0.65\textwidth]{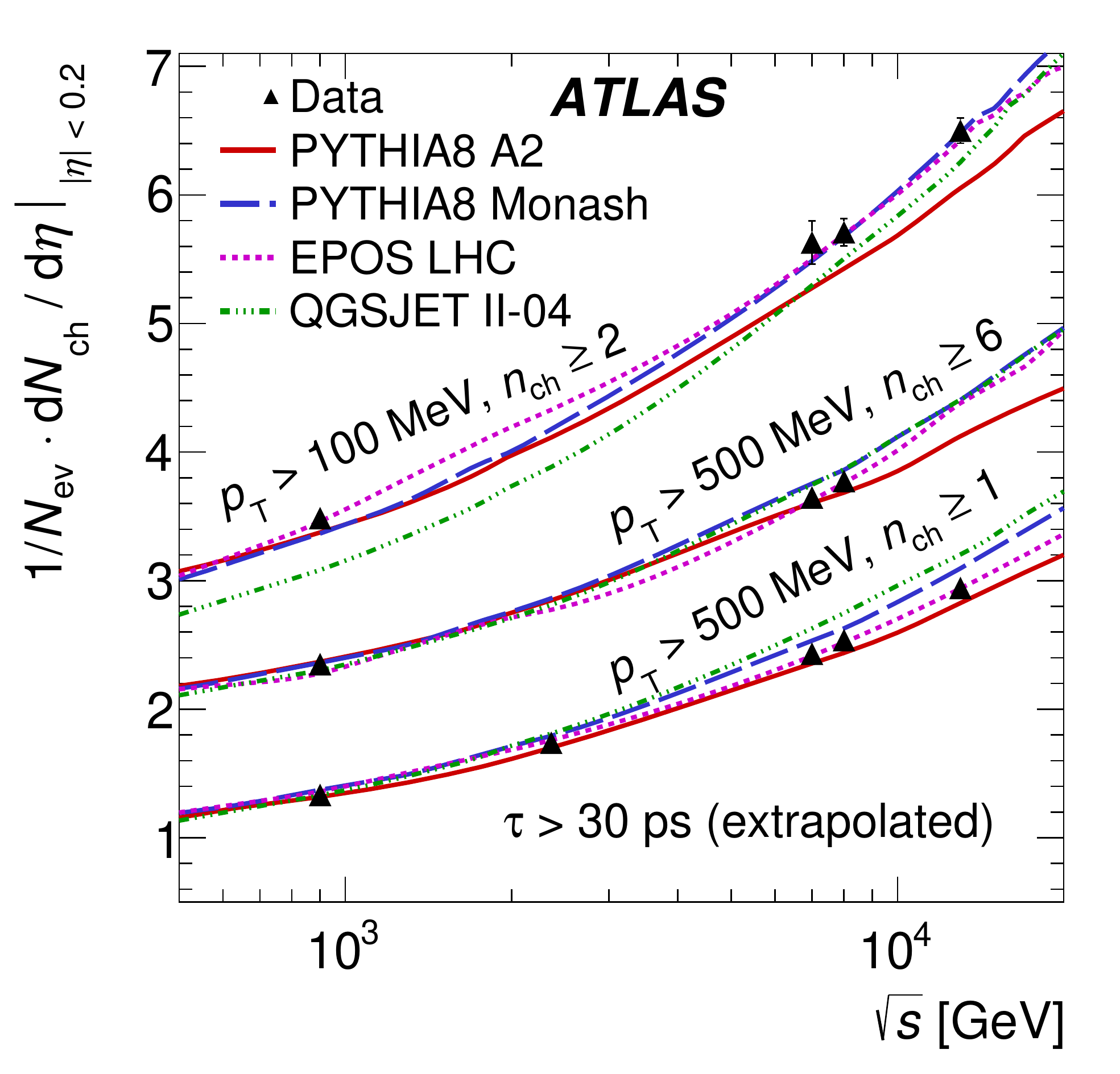}
	\end{center}
	\caption{The average primary charged-particle multiplicity in {\itshape pp} interactions per unit of pseudorapidity \eta for $|\eta| < 0.2$ as a function of the centre-of-mass energy $\sqrt{s}$. The values for the other {\itshape pp} centre-of-mass energies are taken from previous ATLAS analyses~\cite{STDM-2010-06,Paper8TeV}. The value for particles with $\pt>500$\,MeV for a $\sqrt{s}=13$\,TeV is taken from Ref.~\cite{Paper500}. The results have been extrapolated to include charged strange baryons (charged particles with a mean lifetime of $30<\tau<300$\,ps). The data are shown as black triangles with vertical errors bars representing the total uncertainty. They are compared to various MC predictions which are shown as coloured lines. }
	\label{fig:dneta_comparison}
\end{figure*}

The mean number of primary charged particles per unit pseudorapidity in the central \eta region is measured to be $6.422 \pm 0.096$, by averaging over $|\eta| < 0.2$; the quoted error is the systematic uncertainty, the statistical uncertainty is negligible. In order to compare with other measurements, it is corrected for the contribution from strange baryons (and therefore extrapolated to primary charged particles with $\tau > 30$\,ps) by a correction factor of $1.0121 \pm 0.0035$.  The central value is taken from \epos; the systematic uncertainty is taken from the  difference between \epos\ and \py~\atwo\ (the largest difference was observed between \epos\ and \py\ \atwo) and the statistical uncertainty is negligible. The mean number of primary charged particles after the correction is $6.500 \pm 0.099$. This result is compared to previous measurements~\cite{STDM-2010-06,Paper500,Paper8TeV} at different $\sqrt{s}$ values in Figure~\ref{fig:dneta_comparison}. The predictions from \epos\ and \py~\monash\  match the data well. For \py~\atwo, the match is not as good  as was observed when  measuring particles with \pt > 500\,MeV~\cite{Paper500}.

\FloatBarrier

\section{Conclusion}
\label{sec:conclusion}

Primary charged-particle multiplicity measurements with the ATLAS detector using proton--proton collisions delivered by the LHC at $\sqrt{s}=13$\,TeV are presented for events with at least two primary charged particles with $|\eta|<2.5$ and $\pt>100\MeV$ using a specialised track reconstruction algorithm. A data sample corresponding to an integrated luminosity of 151\,$\mu\textrm{b}^{-1}$ is analysed. The mean number of charged particles per unit pseudorapidity in the region $|\eta| < 0.2$ is measured to be $6.422\pm0.096$ with a negligible statistical uncertainty. Significant differences are observed between the measured distributions and the Monte Carlo predictions tested. Amongst the models considered, \epos\ has the best overall description of the data as was seen in a previous ATLAS measurement at $\sqrt{s}=13$\,TeV using tracks with $\pt > 500\MeV$. \py~\atwo\ and \py~\monash\ provide a reasonable overall description, whereas \qgsjet\ does not describe  \meanpt\ vs. \nch\ well but provides a reasonable level of agreement for other distributions.

\section*{Acknowledgements}


%


We thank CERN for the very successful operation of the LHC, as well as the
support staff from our institutions without whom ATLAS could not be
operated efficiently.

We acknowledge the support of ANPCyT, Argentina; YerPhI, Armenia; ARC, Australia; BMWFW and FWF, Austria; ANAS, Azerbaijan; SSTC, Belarus; CNPq and FAPESP, Brazil; NSERC, NRC and CFI, Canada; CERN; CONICYT, Chile; CAS, MOST and NSFC, China; COLCIENCIAS, Colombia; MSMT CR, MPO CR and VSC CR, Czech Republic; DNRF and DNSRC, Denmark; IN2P3-CNRS, CEA-DSM/IRFU, France; GNSF, Georgia; BMBF, HGF, and MPG, Germany; GSRT, Greece; RGC, Hong Kong SAR, China; ISF, I-CORE and Benoziyo Center, Israel; INFN, Italy; MEXT and JSPS, Japan; CNRST, Morocco; FOM and NWO, Netherlands; RCN, Norway; MNiSW and NCN, Poland; FCT, Portugal; MNE/IFA, Romania; MES of Russia and NRC KI, Russian Federation; JINR; MESTD, Serbia; MSSR, Slovakia; ARRS and MIZ\v{S}, Slovenia; DST/NRF, South Africa; MINECO, Spain; SRC and Wallenberg Foundation, Sweden; SERI, SNSF and Cantons of Bern and Geneva, Switzerland; MOST, Taiwan; TAEK, Turkey; STFC, United Kingdom; DOE and NSF, United States of America. In addition, individual groups and members have received support from BCKDF, the Canada Council, CANARIE, CRC, Compute Canada, FQRNT, and the Ontario Innovation Trust, Canada; EPLANET, ERC, FP7, Horizon 2020 and Marie Sk{\l}odowska-Curie Actions, European Union; Investissements d'Avenir Labex and Idex, ANR, R{\'e}gion Auvergne and Fondation Partager le Savoir, France; DFG and AvH Foundation, Germany; Herakleitos, Thales and Aristeia programmes co-financed by EU-ESF and the Greek NSRF; BSF, GIF and Minerva, Israel; BRF, Norway; Generalitat de Catalunya, Generalitat Valenciana, Spain; the Royal Society and Leverhulme Trust, United Kingdom.

The crucial computing support from all WLCG partners is acknowledged gratefully, in particular from CERN, the ATLAS Tier-1 facilities at TRIUMF (Canada), NDGF (Denmark, Norway, Sweden), CC-IN2P3 (France), KIT/GridKA (Germany), INFN-CNAF (Italy), NL-T1 (Netherlands), PIC (Spain), ASGC (Taiwan), RAL (UK) and BNL (USA), the Tier-2 facilities worldwide and large non-WLCG resource providers. Major contributors of computing resources are listed in Ref.~\cite{ATL-GEN-PUB-2016-002}.




\clearpage
\printbibliography
\ifdefined\InternalMode

\clearpage

\appendix
\part*{Auxiliary material}


\input{aux-plots}
\clearpage


\section*{Supplementary material}

# The ATLAS Collaboration

M. Aaboud[135d], G. Aad[86], B. Abbott[113], J. Abdallah[64], O. Abdinov[12], B. Abeloos[117], R. Aben[107],
O.S. AbouZeid[137], N.L. Abraham[149], H. Abramowicz[153], H. Abreu[152], R. Abreu[116], Y. Abulaiti[146a,146b],
B.S. Acharya[163a,163b,a], L. Adamczyk[40a], D.L. Adams[27], J. Adelman[108], S. Adomeit[100], T. Adye[131],
A.A. Affolder[75], T. Agatonovic-Jovin[14], J. Agricola[56], J.A. Aguilar-Saavedra[126a,126f], S.P. Ahlen[24],
F. Ahmadov[66,b], G. Aielli[133a,133b], H. Akerstedt[146a,146b], T.P.A. Åkesson[82], A.V. Akimov[96],
G.L. Alberghi[22a,22b], J. Albert[168], S. Albrand[57], M.J. Alconada Verzini[72], M. Aleksa[32],
I.N. Aleksandrov[66], C. Alexa[28b], G. Alexander[153], T. Alexopoulos[10], M. Alhroob[113], B. Ali[128],
M. Aliev[74a,74b], G. Alimonti[92a], J. Alison[33], S.P. Alkire[37], B.M.M. Allbrooke[149], B.W. Allen[116],
P.P. Allport[19], A. Aloisio[104a,104b], A. Alonso[38], F. Alonso[72], C. Alpigiani[138], M. Alstaty[86],
B. Alvarez Gonzalez[32], D. Álvarez Piqueras[166], M.G. Alviggi[104a,104b], B.T. Amadio[16], K. Amako[67],
Y. Amaral Coutinho[26a], C. Amelung[25], D. Amidei[90], S.P. Amor Dos Santos[126a,126c], A. Amorim[126a,126b],
S. Amoroso[32], G. Amundsen[25], C. Anastopoulos[139], L.S. Ancu[51], N. Andari[108], T. Andeen[11],
C.F. Anders[59b], G. Anders[32], J.K. Anders[75], K.J. Anderson[33], A. Andreazza[92a,92b], V. Andrei[59a],
S. Angelidakis[9], I. Angelozzi[107], P. Anger[46], A. Angerami[37], F. Anghinolfi[32], A.V. Anisenkov[109,c],
N. Anjos[13], A. Annovi[124a,124b], C. Antel[59a], M. Antonelli[49], A. Antonov[98], F. Anulli[132a], M. Aoki[67],
L. Aperio Bella[19], G. Arabidze[91], Y. Arai[67], J.P. Araque[126a], A.T.H. Arce[47], F.A. Arduh[72], J-F. Arguin[95],
S. Argyropoulos[64], M. Arik[20a], A.J. Armbruster[143], L.J. Armitage[77], O. Arnaez[32], H. Arnold[50],
M. Arratia[30], O. Arslan[23], A. Artamonov[97], G. Artoni[120], S. Artz[84], S. Asai[155], N. Asbah[44],
A. Ashkenazi[153], B. Åsman[146a,146b], L. Asquith[149], K. Assamagan[27], R. Astalos[144a], M. Atkinson[165],
N.B. Atlay[141], K. Augsten[128], G. Avolio[32], B. Axen[16], M.K. Ayoub[117], G. Azuelos[95,d], M.A. Baak[32],
A.E. Baas[59a], M.J. Baca[19], H. Bachacou[136], K. Bachas[74a,74b], M. Backes[32], M. Backhaus[32],
P. Bagiacchi[132a,132b], P. Bagnaia[132a,132b], Y. Bai[35a], J.T. Baines[131], O.K. Baker[175], E.M. Baldin[109,c],
P. Balek[171], T. Balestri[148], F. Balli[136], W.K. Balunas[122], E. Banas[41], Sw. Banerjee[172,e],
A.A.E. Bannoura[174], L. Barak[32], E.L. Barberio[89], D. Barberis[52a,52b], M. Barbero[86], T. Barillari[101],
M-S Barisits[32], T. Barklow[143], N. Barlow[30], S.L. Barnes[85], B.M. Barnett[131], R.M. Barnett[16],
Z. Barnovska[5], A. Baroncelli[134a], G. Barone[25], A.J. Barr[120], L. Barranco Navarro[166], F. Barreiro[83],
J. Barreiro Guimarães da Costa[35a], R. Bartoldus[143], A.E. Barton[73], P. Bartos[144a], A. Basalaev[123],
A. Bassalat[117], R.L. Bates[55], S.J. Batista[158], J.R. Batley[30], M. Battaglia[137], M. Bauce[132a,132b],
F. Bauer[136], H.S. Bawa[143,f], J.B. Beacham[111], M.D. Beattie[73], T. Beau[81], P.H. Beauchemin[161],
P. Bechtle[23], H.P. Beck[18,g], K. Becker[120], M. Becker[84], M. Beckingham[169], C. Becot[110], A.J. Beddall[20e],
A. Beddall[20b], V.A. Bednyakov[66], M. Bedognetti[107], C.P. Bee[148], L.J. Beemster[107], T.A. Beermann[32],
M. Begel[27], J.K. Behr[44], C. Belanger-Champagne[88], A.S. Bell[79], G. Bella[153], L. Bellagamba[22a],
A. Bellerive[31], M. Bellomo[87], K. Belotskiy[98], O. Beltramello[32], N.L. Belyaev[98], O. Benary[153],
D. Benchekroun[135a], M. Bender[100], K. Bendtz[146a,146b], N. Benekos[10], Y. Benhammou[153],
E. Benhar Noccioli[175], J. Benitez[64], D.P. Benjamin[47], J.R. Bensinger[25], S. Bentvelsen[107],
L. Beresford[120], M. Beretta[49], D. Berge[107], E. Bergeaas Kuutmann[164], N. Berger[5], J. Beringer[16],
S. Berlendis[57], N.R. Bernard[87], C. Bernius[110], F.U. Bernlochner[23], T. Berry[78], P. Berta[129], C. Bertella[84],
G. Bertoli[146a,146b], F. Bertolucci[124a,124b], I.A. Bertram[73], C. Bertsche[44], D. Bertsche[113], G.J. Besjes[38],
O. Bessidskaia Bylund[146a,146b], M. Bessner[44], N. Besson[136], C. Betancourt[50], S. Bethke[101],
A.J. Bevan[77], W. Bhimji[16], R.M. Bianchi[125], L. Bianchini[25], M. Bianco[32], O. Biebel[100],
D. Biedermann[17], R. Bielski[85], N.V. Biesuz[124a,124b], M. Biglietti[134a], J. Bilbao De Mendizabal[51],
H. Bilokon[49], M. Bindi[56], S. Binet[117], A. Bingul[20b], C. Bini[132a,132b], S. Biondi[22a,22b], D.M. Bjergaard[47],
C.W. Black[150], J.E. Black[143], K.M. Black[24], D. Blackburn[138], R.E. Blair[6], J.-B. Blanchard[136],
J.E. Blanco[78], T. Blazek[144a], I. Bloch[44], C. Blocker[25], W. Blum[84,*], U. Blumenschein[56], S. Blunier[34a],

G.J. Bobbink[107], V.S. Bobrovnikov[109,c], S.S. Bocchetta[82], A. Bocci[47], C. Bock[100], M. Boehler[50],
D. Boerner[174], J.A. Bogaerts[32], D. Bogavac[14], A.G. Bogdanchikov[109], C. Bohm[146a], V. Boisvert[78],
P. Bokan[14], T. Bold[40a], A.S. Boldyrev[163a,163c], M. Bomben[81], M. Bona[77], M. Boonekamp[136],
A. Borisov[130], G. Borissov[73], J. Bortfeldt[32], D. Bortoletto[120], V. Bortolotto[61a,61b,61c], K. Bos[107],
D. Boscherini[22a], M. Bosman[13], J.D. Bossio Sola[29], J. Boudreau[125], J. Bouffard[2],
E.V. Bouhova-Thacker[73], D. Boumediene[36], C. Bourdarios[117], S.K. Boutle[55], A. Boveia[32], J. Boyd[32],
I.R. Boyko[66], J. Bracinik[19], A. Brandt[8], G. Brandt[56], O. Brandt[59a], U. Bratzler[156], B. Brau[87],
J.E. Brau[116], H.M. Braun[174,*], W.D. Breaden Madden[55], K. Brendlinger[122], A.J. Brennan[89],
L. Brenner[107], R. Brenner[164], S. Bressler[171], T.M. Bristow[48], D. Britton[55], D. Britzger[44], F.M. Brochu[30],
I. Brock[23], R. Brock[91], G. Brooijmans[37], T. Brooks[78], W.K. Brooks[34b], J. Brosamer[16], E. Brost[116],
J.H Broughton[19], P.A. Bruckman de Renstrom[41], D. Bruncko[144b], R. Bruneliere[50], A. Bruni[22a],
G. Bruni[22a], L.S. Bruni[107], BH Brunt[30], M. Bruschi[22a], N. Bruscino[23], P. Bryant[33], L. Bryngemark[82],
T. Buanes[15], Q. Buat[142], P. Buchholz[141], A.G. Buckley[55], I.A. Budagov[66], F. Buehrer[50], M.K. Bugge[119],
O. Bulekov[98], D. Bullock[8], H. Burckhart[32], S. Burdin[75], C.D. Burgard[50], B. Burghgrave[108], K. Burka[41],
S. Burke[131], I. Burmeister[45], J.T.P. Burr[120], E. Busato[36], D. Büscher[50], V. Büscher[84], P. Bussey[55],
J.M. Butler[24], C.M. Buttar[55], J.M. Butterworth[79], P. Butti[107], W. Buttinger[27], A. Buzatu[55],
A.R. Buzykaev[109,c], S. Cabrera Urbán[166], D. Caforio[128], V.M. Cairo[39a,39b], O. Cakir[4a], N. Calace[51],
P. Calafiura[16], A. Calandri[86], G. Calderini[81], P. Calfayan[100], G. Callea[39a,39b], L.P. Caloba[26a],
S. Calvente Lopez[83], D. Calvet[36], S. Calvet[36], T.P. Calvet[86], R. Camacho Toro[33], S. Camarda[32],
P. Camarri[133a,133b], D. Cameron[119], R. Caminal Armadans[165], C. Camincher[57], S. Campana[32],
M. Campanelli[79], A. Camplani[92a,92b], A. Campoverde[141], V. Canale[104a,104b], A. Canepa[159a],
M. Cano Bret[35e], J. Cantero[114], R. Cantrill[126a], T. Cao[42], M.D.M. Capeans Garrido[32], I. Caprini[28b],
M. Caprini[28b], M. Capua[39a,39b], R. Caputo[84], R.M. Carbone[37], R. Cardarelli[133a], F. Cardillo[50],
I. Carli[129], T. Carli[32], G. Carlino[104a], L. Carminati[92a,92b], S. Caron[106], E. Carquin[34b],
G.D. Carrillo-Montoya[32], J.R. Carter[30], J. Carvalho[126a,126c], D. Casadei[19], M.P. Casado[13,h],
M. Casolino[13], D.W. Casper[162], E. Castaneda-Miranda[145a], R. Castelijn[107], A. Castelli[107],
V. Castillo Gimenez[166], N.F. Castro[126a,i], A. Catinaccio[32], J.R. Catmore[119], A. Cattai[32], J. Caudron[84],
V. Cavaliere[165], E. Cavallaro[13], D. Cavalli[92a], M. Cavalli-Sforza[13], V. Cavasinni[124a,124b],
F. Ceradini[134a,134b], L. Cerda Alberich[166], B.C. Cerio[47], A.S. Cerqueira[26b], A. Cerri[149], L. Cerrito[77],
F. Cerutti[16], M. Cerv[32], A. Cervelli[18], S.A. Cetin[20d], A. Chafaq[135a], D. Chakraborty[108], S.K. Chan[58],
Y.L. Chan[61a], P. Chang[165], J.D. Chapman[30], D.G. Charlton[19], A. Chatterjee[51], C.C. Chau[158],
C.A. Chavez Barajas[149], S. Che[111], S. Cheatham[73], A. Chegwidden[91], S. Chekanov[6],
S.V. Chekulaev[159a], G.A. Chelkov[66,j], M.A. Chelstowska[90], C. Chen[65], H. Chen[27], K. Chen[148],
S. Chen[35c], S. Chen[155], X. Chen[35f], Y. Chen[68], H.C. Cheng[90], H.J Cheng[35a], Y. Cheng[33],
A. Cheplakov[66], E. Cheremushkina[130], R. Cherkaoui El Moursli[135e], V. Chernyatin[27,*], E. Cheu[7],
L. Chevalier[136], V. Chiarella[49], G. Chiarelli[124a,124b], G. Chiodini[74a], A.S. Chisholm[19], A. Chitan[28b],
M.V. Chizhov[66], K. Choi[62], A.R. Chomont[36], S. Chouridou[9], B.K.B. Chow[100], V. Christodoulou[79],
D. Chromek-Burckhart[32], J. Chudoba[127], A.J. Chuinard[88], J.J. Chwastowski[41], L. Chytka[115],
G. Ciapetti[132a,132b], A.K. Ciftci[4a], D. Cinca[45], V. Cindro[76], I.A. Cioara[23], C. Ciocca[22a,22b], A. Ciocio[16],
F. Cirotto[104a,104b], Z.H. Citron[171], M. Citterio[92a], M. Ciubancan[28b], A. Clark[51], B.L. Clark[58],
M.R. Clark[37], P.J. Clark[48], R.N. Clarke[16], C. Clement[146a,146b], Y. Coadou[86], M. Cobal[163a,163c],
A. Coccaro[51], J. Cochran[65], L. Coffey[25], L. Colasurdo[106], B. Cole[37], A.P. Colijn[107], J. Collot[57],
T. Colombo[32], G. Compostella[101], P. Conde Muiño[126a,126b], E. Coniavitis[50], S.H. Connell[145b],
I.A. Connelly[78], V. Consorti[50], S. Constantinescu[28b], G. Conti[32], F. Conventi[104a,k], M. Cooke[16],
B.D. Cooper[79], A.M. Cooper-Sarkar[120], K.J.R. Cormier[158], T. Cornelissen[174], M. Corradi[132a,132b],
F. Corriveau[88,l], A. Corso-Radu[162], A. Cortes-Gonzalez[13], G. Cortiana[101], G. Costa[92a], M.J. Costa[166],
D. Costanzo[139], G. Cottin[30], G. Cowan[78], B.E. Cox[85], K. Cranmer[110], S.J. Crawley[55], G. Cree[31],

S. Crépé-Renaudin[57], F. Crescioli[81], W.A. Cribbs[146a,146b], M. Crispin Ortuzar[120], M. Cristinziani[23],
V. Croft[106], G. Crosetti[39a,39b], T. Cuhadar Donszelmann[139], J. Cummings[175], M. Curatolo[49], J. Cúth[84],
C. Cuthbert[150], H. Czirr[141], P. Czodrowski[3], G. D'amen[22a,22b], S. D'Auria[55], M. D'Onofrio[75],
M.J. Da Cunha Sargedas De Sousa[126a,126b], C. Da Via[85], W. Dabrowski[40a], T. Dado[144a], T. Dai[90],
O. Dale[15], F. Dallaire[95], C. Dallapiccola[87], M. Dam[38], J.R. Dandoy[33], N.P. Dang[50], A.C. Daniells[19],
N.S. Dann[85], M. Danninger[167], M. Dano Hoffmann[136], V. Dao[50], G. Darbo[52a], S. Darmora[8],
J. Dassoulas[3], A. Dattagupta[62], W. Davey[23], C. David[168], T. Davidek[129], M. Davies[153], P. Davison[79],
E. Dawe[89], I. Dawson[139], R.K. Daya-Ishmukhametova[87], K. De[8], R. de Asmundis[104a],
A. De Benedetti[113], S. De Castro[22a,22b], S. De Cecco[81], N. De Groot[106], P. de Jong[107], H. De la Torre[83],
F. De Lorenzi[65], A. De Maria[56], D. De Pedis[132a], A. De Salvo[132a], U. De Sanctis[149], A. De Santo[149],
J.B. De Vivie De Regie[117], W.J. Dearnaley[73], R. Debbe[27], C. Debenedetti[137], D.V. Dedovich[66],
N. Dehghanian[3], I. Deigaard[107], M. Del Gaudio[39a,39b], J. Del Peso[83], T. Del Prete[124a,124b],
D. Delgove[117], F. Deliot[136], C.M. Delitzsch[51], M. Deliyergiyev[76], A. Dell'Acqua[32], L. Dell'Asta[24],
M. Dell'Orso[124a,124b], M. Della Pietra[104a,k], D. della Volpe[51], M. Delmastro[5], P.A. Delsart[57],
D.A. DeMarco[158], S. Demers[175], M. Demichev[66], A. Demilly[81], S.P. Denisov[130], D. Denysiuk[136],
D. Derendarz[41], J.E. Derkaoui[135d], F. Derue[81], P. Dervan[75], K. Desch[23], C. Deterre[44], K. Dette[45],
M.R. Devesa[29], P.O. Deviveiros[32], A. Dewhurst[131], S. Dhaliwal[25], A. Di Ciaccio[133a,133b], L. Di Ciaccio[5],
W.K. Di Clemente[122], C. Di Donato[132a,132b], A. Di Girolamo[32], B. Di Girolamo[32], B. Di Micco[134a,134b],
R. Di Nardo[32], A. Di Simone[50], R. Di Sipio[158], D. Di Valentino[31], C. Diaconu[86], M. Diamond[158],
F.A. Dias[48], M.A. Diaz[34a], E.B. Diehl[90], J. Dietrich[17], S. Diglio[86], A. Dimitrievska[14], J. Dingfelder[23],
P. Dita[28b], S. Dita[28b], F. Dittus[32], F. Djama[86], T. Djobava[53b], J.I. Djuvsland[59a], M.A.B. do Vale[26c],
D. Dobos[32], M. Dobre[28b], C. Doglioni[82], T. Dohmae[155], J. Dolejsi[129], Z. Dolezal[129],
B.A. Dolgoshein[98,*], M. Donadelli[26d], S. Donati[124a,124b], P. Dondero[121a,121b], J. Donini[36], J. Dopke[131],
A. Doria[104a], M.T. Dova[72], A.T. Doyle[55], E. Drechsler[56], M. Dris[10], Y. Du[35d], J. Duarte-Campderros[153],
E. Duchovni[171], G. Duckeck[100], O.A. Ducu[95,m], D. Duda[107], A. Dudarev[32], E.M. Duffield[16],
L. Duflot[117], L. Duguid[78], M. Dührssen[32], M. Dumancic[171], M. Dunford[59a], H. Duran Yildiz[4a],
M. Düren[54], A. Durglishvili[53b], D. Duschinger[46], B. Dutta[44], M. Dyndal[44], C. Eckardt[44], K.M. Ecker[101],
R.C. Edgar[90], N.C. Edwards[48], T. Eifert[32], G. Eigen[15], K. Einsweiler[16], T. Ekelof[164], M. El Kacimi[135c],
V. Ellajosyula[86], M. Ellert[164], S. Elles[5], F. Ellinghaus[174], A.A. Elliot[168], N. Ellis[32], J. Elmsheuser[27],
M. Elsing[32], D. Emeliyanov[131], Y. Enari[155], O.C. Endner[84], M. Endo[118], J.S. Ennis[169], J. Erdmann[45],
A. Ereditato[18], G. Ernis[174], J. Ernst[2], M. Ernst[27], S. Errede[165], E. Ertel[84], M. Escalier[117], H. Esch[45],
C. Escobar[125], B. Esposito[49], A.I. Etienvre[136], E. Etzion[153], H. Evans[62], A. Ezhilov[123], F. Fabbri[22a,22b],
L. Fabbri[22a,22b], G. Facini[33], R.M. Fakhrutdinov[130], S. Falciano[132a], R.J. Falla[79], J. Faltova[129],
Y. Fang[35a], M. Fanti[92a,92b], A. Farbin[8], A. Farilla[134a], C. Farina[125], E.M. Farina[121a,121b], T. Farooque[13],
S. Farrell[16], S.M. Farrington[169], P. Farthouat[32], F. Fassi[135e], P. Fassnacht[32], D. Fassouliotis[9],
M. Faucci Giannelli[78], A. Favareto[52a,52b], W.J. Fawcett[120], L. Fayard[117], O.L. Fedin[123,n], W. Fedorko[167],
S. Feigl[119], L. Feligioni[86], C. Feng[35d], E.J. Feng[32], H. Feng[90], A.B. Fenyuk[130], L. Feremenga[8],
P. Fernandez Martinez[166], S. Fernandez Perez[13], J. Ferrando[55], A. Ferrari[164], P. Ferrari[107], R. Ferrari[121a],
D.E. Ferreira de Lima[59b], A. Ferrer[166], D. Ferrere[51], C. Ferretti[90], A. Ferretto Parodi[52a,52b], F. Fiedler[84],
A. Filipčič[76], M. Filipuzzi[44], F. Filthaut[106], M. Fincke-Keeler[168], K.D. Finelli[150],
M.C.N. Fiolhais[126a,126c], L. Fiorini[166], A. Firan[42], A. Fischer[2], C. Fischer[13], J. Fischer[174], W.C. Fisher[91],
N. Flaschel[44], I. Fleck[141], P. Fleischmann[90], G.T. Fletcher[139], R.R.M. Fletcher[122], T. Flick[174],
A. Floderus[82], L.R. Flores Castillo[61a], M.J. Flowerdew[101], G.T. Forcolin[85], A. Formica[136], A. Forti[85],
A.G. Foster[19], D. Fournier[117], H. Fox[73], S. Fracchia[13], P. Francavilla[81], M. Franchini[22a,22b],
D. Francis[32], L. Franconi[119], M. Franklin[58], M. Frate[162], M. Fraternali[121a,121b], D. Freeborn[79],
S.M. Fressard-Batraneanu[32], F. Friedrich[46], D. Froidevaux[32], J.A. Frost[120], C. Fukunaga[156],
E. Fullana Torregrosa[84], T. Fusayasu[102], J. Fuster[166], C. Gabaldon[57], O. Gabizon[174], A. Gabrielli[22a,22b],

A. Gabrielli[16], G.P. Gach[40a], S. Gadatsch[32], S. Gadomski[51], G. Gagliardi[52a,52b], L.G. Gagnon[95],
P. Gagnon[62], C. Galea[106], B. Galhardo[126a,126c], E.J. Gallas[120], B.J. Gallop[131], P. Gallus[128], G. Galster[38],
K.K. Gan[111], J. Gao[35b,86], Y. Gao[48], Y.S. Gao[143,f], F.M. Garay Walls[48], C. García[166],
J.E. García Navarro[166], M. Garcia-Sciveres[16], R.W. Gardner[33], N. Garelli[143], V. Garonne[119],
A. Gascon Bravo[44], C. Gatti[49], A. Gaudiello[52a,52b], G. Gaudio[121a], B. Gaur[141], L. Gauthier[95],
I.L. Gavrilenko[96], C. Gay[167], G. Gaycken[23], E.N. Gazis[10], Z. Gecse[167], C.N.P. Gee[131],
Ch. Geich-Gimbel[23], M. Geisen[84], M.P. Geisler[59a], C. Gemme[52a], M.H. Genest[57], C. Geng[35b,o],
S. Gentile[132a,132b], C. Gentsos[154], S. George[78], D. Gerbaudo[13], A. Gershon[153], S. Ghasemi[141],
H. Ghazlane[135b], M. Ghneimat[23], B. Giacobbe[22a], S. Giagu[132a,132b], P. Giannetti[124a,124b], B. Gibbard[27],
S.M. Gibson[78], M. Gignac[167], M. Gilchriese[16], T.P.S. Gillam[30], D. Gillberg[31], G. Gilles[174],
D.M. Gingrich[3,d], N. Giokaris[9], M.P. Giordani[163a,163c], F.M. Giorgi[22a], F.M. Giorgi[17], P.F. Giraud[136],
P. Giromini[58], D. Giugni[92a], F. Giuli[120], C. Giuliani[101], M. Giulini[59b], B.K. Gjelsten[119], S. Gkaitatzis[154],
I. Gkialas[154], E.L. Gkougkousis[117], L.K. Gladilin[99], C. Glasman[83], J. Glatzer[32], P.C.F. Glaysher[48],
A. Glazov[44], M. Goblirsch-Kolb[25], J. Godlewski[41], S. Goldfarb[89], T. Golling[51], D. Golubkov[130],
A. Gomes[126a,126b,126d], R. Gonçalo[126a], J. Goncalves Pinto Firmino Da Costa[136], G. Gonella[50],
L. Gonella[19], A. Gongadze[66], S. González de la Hoz[166], G. Gonzalez Parra[13], S. Gonzalez-Sevilla[51],
L. Goossens[32], P.A. Gorbounov[97], H.A. Gordon[27], I. Gorelov[105], B. Gorini[32], E. Gorini[74a,74b],
A. Gorišek[76], E. Gornicki[41], A.T. Goshaw[47], C. Gössling[45], M.I. Gostkin[66], C.R. Goudet[117],
D. Goujdami[135c], A.G. Goussiou[138], N. Govender[145b,p], E. Gozani[152], L. Graber[56],
I. Grabowska-Bold[40a], P.O.J. Gradin[57], P. Grafström[22a,22b], J. Gramling[51], E. Gramstad[119],
S. Grancagnolo[17], V. Gratchev[123], P.M. Gravila[28e], H.M. Gray[32], E. Graziani[134a], Z.D. Greenwood[80,q],
C. Grefe[23], K. Gregersen[79], I.M. Gregor[44], P. Grenier[143], K. Grevtsov[5], J. Griffiths[8], A.A. Grillo[137],
K. Grimm[73], S. Grinstein[13,r], Ph. Gris[36], J.-F. Grivaz[117], S. Groh[84], J.P. Grohs[46], E. Gross[171],
J. Grosse-Knetter[56], G.C. Grossi[80], Z.J. Grout[149], L. Guan[90], W. Guan[172], J. Guenther[63], F. Guescini[51],
D. Guest[162], O. Gueta[153], E. Guido[52a,52b], T. Guillemin[5], S. Guindon[2], U. Gul[55], C. Gumpert[32],
J. Guo[35e], Y. Guo[35b,o], R. Gupta[42], S. Gupta[120], G. Gustavino[132a,132b], P. Gutierrez[113],
N.G. Gutierrez Ortiz[79], C. Gutschow[46], C. Guyot[136], C. Gwenlan[120], C.B. Gwilliam[75], A. Haas[110],
C. Haber[16], H.K. Hadavand[8], N. Haddad[135e], A. Hadef[86], P. Haefner[23], S. Hageböck[23], Z. Hajduk[41],
H. Hakobyan[176,*], M. Haleem[44], J. Haley[114], G. Halladjian[91], G.D. Hallewell[86], K. Hamacher[174],
P. Hamal[115], K. Hamano[168], A. Hamilton[145a], G.N. Hamity[139], P.G. Hamnett[44], L. Han[35b],
K. Hanagaki[67,s], K. Hanawa[155], M. Hance[137], B. Haney[122], S. Hanisch[32], P. Hanke[59a], R. Hanna[136],
J.B. Hansen[38], J.D. Hansen[38], M.C. Hansen[23], P.H. Hansen[38], K. Hara[160], A.S. Hard[172],
T. Harenberg[174], F. Hariri[117], S. Harkusha[93], R.D. Harrington[48], P.F. Harrison[169], F. Hartjes[107],
N.M. Hartmann[100], M. Hasegawa[68], Y. Hasegawa[140], A. Hasib[113], S. Hassani[136], S. Haug[18],
R. Hauser[91], L. Hauswald[46], M. Havranek[127], C.M. Hawkes[19], R.J. Hawkings[32], D. Hayden[91],
C.P. Hays[120], J.M. Hays[77], H.S. Hayward[75], S.J. Haywood[131], S.J. Head[19], T. Heck[84], V. Hedberg[82],
L. Heelan[8], S. Heim[122], T. Heim[16], B. Heinemann[16], J.J. Heinrich[100], L. Heinrich[110], C. Heinz[54],
J. Hejbal[127], L. Helary[24], S. Hellman[146a,146b], C. Helsens[32], J. Henderson[120], R.C.W. Henderson[73],
Y. Heng[172], S. Henkelmann[167], A.M. Henriques Correia[32], S. Henrot-Versille[117], G.H. Herbert[17],
Y. Hernández Jiménez[166], G. Herten[50], R. Hertenberger[100], L. Hervas[32], G.G. Hesketh[79], N.P. Hessey[107],
J.W. Hetherly[42], R. Hickling[77], E. Higón-Rodriguez[166], E. Hill[168], J.C. Hill[30], K.H. Hiller[44],
S.J. Hillier[19], I. Hinchliffe[16], E. Hines[122], R.R. Hinman[16], M. Hirose[50], D. Hirschbuehl[174], J. Hobbs[148],
N. Hod[159a], M.C. Hodgkinson[139], P. Hodgson[139], A. Hoecker[32], M.R. Hoeferkamp[105], F. Hoenig[100],
D. Hohn[23], T.R. Holmes[16], M. Homann[45], T.M. Hong[125], B.H. Hooberman[165], W.H. Hopkins[116],
Y. Horii[103], A.J. Horton[142], J-Y. Hostachy[57], S. Hou[151], A. Hoummada[135a], J. Howarth[44],
M. Hrabovsky[115], I. Hristova[17], J. Hrivnac[117], T. Hryn'ova[5], A. Hrynevich[94], C. Hsu[145c], P.J. Hsu[151,t],
S.-C. Hsu[138], D. Hu[37], Q. Hu[35b], Y. Huang[44], Z. Hubacek[128], F. Hubaut[86], F. Huegging[23],

T.B. Huffman[120], E.W. Hughes[37], G. Hughes[73], M. Huhtinen[32], P. Huo[148], N. Huseynov[66,b], J. Huston[91],
J. Huth[58], G. Iacobucci[51], G. Iakovidis[27], I. Ibragimov[141], L. Iconomidou-Fayard[117], E. Ideal[175],
Z. Idrissi[135e], P. Iengo[32], O. Igonkina[107,u], T. Iizawa[170], Y. Ikegami[67], M. Ikeno[67], Y. Ilchenko[11,v],
D. Iliadis[154], N. Ilic[143], T. Ince[101], G. Introzzi[121a,121b], P. Ioannou[9,*], M. Iodice[134a], K. Iordanidou[37],
V. Ippolito[58], N. Ishijima[118], M. Ishino[69], M. Ishitsuka[157], R. Ishmukhametov[111], C. Issever[120],
S. Istin[20a], F. Ito[160], J.M. Iturbe Ponce[85], R. Iuppa[133a,133b], W. Iwanski[41], H. Iwasaki[67], J.M. Izen[43],
V. Izzo[104a], S. Jabbar[3], B. Jackson[122], M. Jackson[75], P. Jackson[1], V. Jain[2], K.B. Jakobi[84], K. Jakobs[50],
S. Jakobsen[32], T. Jakoubek[127], D.O. Jamin[114], D.K. Jana[80], E. Jansen[79], R. Jansky[63], J. Janssen[23],
M. Janus[56], G. Jarlskog[82], N. Javadov[66,b], T. Javůrek[50], F. Jeanneau[136], L. Jeanty[16], J. Jejelava[53a,w],
G.-Y. Jeng[150], D. Jennens[89], P. Jenni[50,x], J. Jentzsch[45], C. Jeske[169], S. Jézéquel[5], H. Ji[172], J. Jia[148],
H. Jiang[65], Y. Jiang[35b], S. Jiggins[79], J. Jimenez Pena[166], S. Jin[35a], A. Jinaru[28b], O. Jinnouchi[157],
P. Johansson[139], K.A. Johns[7], W.J. Johnson[138], K. Jon-And[146a,146b], G. Jones[169], R.W.L. Jones[73],
S. Jones[7], T.J. Jones[75], J. Jongmanns[59a], P.M. Jorge[126a,126b], J. Jovicevic[159a], X. Ju[172],
A. Juste Rozas[13,r], M.K. Köhler[171], A. Kaczmarska[41], M. Kado[117], H. Kagan[111], M. Kagan[143],
S.J. Kahn[86], E. Kajomovitz[47], C.W. Kalderon[120], A. Kaluza[84], S. Kama[42], A. Kamenshchikov[130],
N. Kanaya[155], S. Kaneti[30], L. Kanjir[76], V.A. Kantserov[98], J. Kanzaki[67], B. Kaplan[110], L.S. Kaplan[172],
A. Kapliy[33], D. Kar[145c], K. Karakostas[10], A. Karamaoun[3], N. Karastathis[10], M.J. Kareem[56],
E. Karentzos[10], M. Karnevskiy[84], S.N. Karpov[66], Z.M. Karpova[66], K. Karthik[110], V. Kartvelishvili[73],
A.N. Karyukhin[130], K. Kasahara[160], L. Kashif[172], R.D. Kass[111], A. Kastanas[15], Y. Kataoka[155],
C. Kato[155], A. Katre[51], J. Katzy[44], K. Kawagoe[71], T. Kawamoto[155], G. Kawamura[56], S. Kazama[155],
V.F. Kazanin[109,c], R. Keeler[168], R. Kehoe[42], J.S. Keller[44], J.J. Kempster[78], K Kentaro[103],
H. Keoshkerian[158], O. Kepka[127], B.P. Kerševan[76], S. Kersten[174], R.A. Keyes[88], M. Khader[165],
F. Khalil-zada[12], A. Khanov[114], A.G. Kharlamov[109,c], T.J. Khoo[51], V. Khovanskiy[97], E. Khramov[66],
J. Khubua[53b,y], S. Kido[68], H.Y. Kim[8], S.H. Kim[160], Y.K. Kim[33], N. Kimura[154], O.M. Kind[17],
B.T. King[75], M. King[166], S.B. King[167], J. Kirk[131], A.E. Kiryunin[101], T. Kishimoto[68], D. Kisielewska[40a],
F. Kiss[50], K. Kiuchi[160], O. Kivernyk[136], E. Kladiva[144b], M.H. Klein[37], M. Klein[75], U. Klein[75],
K. Kleinknecht[84], P. Klimek[108], A. Klimentov[27], R. Klingenberg[45], J.A. Klinger[139], T. Klioutchnikova[32],
E.-E. Kluge[59a], P. Kluit[107], S. Kluth[101], J. Knapik[41], E. Kneringer[63], E.B.F.G. Knoops[86], A. Knue[55],
A. Kobayashi[155], D. Kobayashi[157], T. Kobayashi[155], M. Kobel[46], M. Kocian[143], P. Kodys[129], T. Koffas[31],
E. Koffeman[107], T. Koi[143], H. Kolanoski[17], M. Kolb[59b], I. Koletsou[5], A.A. Komar[96,*], Y. Komori[155],
T. Kondo[67], N. Kondrashova[44], K. Köneke[50], A.C. König[106], T. Kono[67,z], R. Konoplich[110,aa],
N. Konstantinidis[79], R. Kopeliansky[62], S. Koperny[40a], L. Köpke[84], A.K. Kopp[50], K. Korcyl[41],
K. Kordas[154], A. Korn[79], A.A. Korol[109,c], I. Korolkov[13], E.V. Korolkova[139], O. Kortner[101],
S. Kortner[101], T. Kosek[129], V.V. Kostyukhin[23], A. Kotwal[47], A. Kourkoumeli-Charalampidi[154],
C. Kourkoumelis[9], V. Kouskoura[27], A.B. Kowalewska[41], R. Kowalewski[168], T.Z. Kowalski[40a],
C. Kozakai[155], W. Kozanecki[136], A.S. Kozhin[130], V.A. Kramarenko[99], G. Kramberger[76],
D. Krasnopevtsev[98], M.W. Krasny[81], A. Krasznahorkay[32], J.K. Kraus[23], A. Kravchenko[27], M. Kretz[59c],
J. Kretzschmar[75], K. Kreutzfeldt[54], P. Krieger[158], K. Krizka[33], K. Kroeninger[45], H. Kroha[101],
J. Kroll[122], J. Kroseberg[23], J. Krstic[14], U. Kruchonak[66], H. Krüger[23], N. Krumnack[65], A. Kruse[172],
M.C. Kruse[47], M. Kruskal[24], T. Kubota[89], H. Kucuk[79], S. Kuday[4b], J.T. Kuechler[174], S. Kuehn[50],
A. Kugel[59c], F. Kuger[173], A. Kuhl[137], T. Kuhl[44], V. Kukhtin[66], R. Kukla[136], Y. Kulchitsky[93],
S. Kuleshov[34b], M. Kuna[132a,132b], T. Kunigo[69], A. Kupco[127], H. Kurashige[68], Y.A. Kurochkin[93],
V. Kus[127], E.S. Kuwertz[168], M. Kuze[157], J. Kvita[115], T. Kwan[168], D. Kyriazopoulos[139], A. La Rosa[101],
J.L. La Rosa Navarro[26d], L. La Rotonda[39a,39b], C. Lacasta[166], F. Lacava[132a,132b], J. Lacey[31], H. Lacker[17],
D. Lacour[81], V.R. Lacuesta[166], E. Ladygin[66], R. Lafaye[5], B. Laforge[81], T. Lagouri[175], S. Lai[56],
S. Lammers[62], W. Lampl[7], E. Lançon[136], U. Landgraf[50], M.P.J. Landon[77], M.C. Lanfermann[51],
V.S. Lang[59a], J.C. Lange[13], A.J. Lankford[162], F. Lanni[27], K. Lantzsch[23], A. Lanza[121a], S. Laplace[81],

C. Lapoire[32], J.F. Laporte[136], T. Lari[92a], F. Lasagni Manghi[22a,22b], M. Lassnig[32], P. Laurelli[49],
W. Lavrijsen[16], A.T. Law[137], P. Laycock[75], T. Lazovich[58], M. Lazzaroni[92a,92b], B. Le[89], O. Le Dortz[81],
E. Le Guirriec[86], E.P. Le Quilleuc[136], M. LeBlanc[168], T. LeCompte[6], F. Ledroit-Guillon[57], C.A. Lee[27],
S.C. Lee[151], L. Lee[1], G. Lefebvre[81], M. Lefebvre[168], F. Legger[100], C. Leggett[16], A. Lehan[75],
G. Lehmann Miotto[32], X. Lei[7], W.A. Leight[31], A. Leisos[154,ab], A.G. Leister[175], M.A.L. Leite[26d],
R. Leitner[129], D. Lellouch[171], B. Lemmer[56], K.J.C. Leney[79], T. Lenz[23], B. Lenzi[32], R. Leone[7],
S. Leone[124a,124b], C. Leonidopoulos[48], S. Leontsinis[10], G. Lerner[149], C. Leroy[95], A.A.J. Lesage[136],
C.G. Lester[30], M. Levchenko[123], J. Levêque[5], D. Levin[90], L.J. Levinson[171], M. Levy[19], D. Lewis[77],
A.M. Leyko[23], M. Leyton[43], B. Li[35b,o], H. Li[148], H.L. Li[33], L. Li[47], L. Li[35e], Q. Li[35a], S. Li[47], X. Li[85],
Y. Li[141], Z. Liang[35a], B. Liberti[133a], A. Liblong[158], P. Lichard[32], K. Lie[165], J. Liebal[23], W. Liebig[15],
A. Limosani[150], S.C. Lin[151,ac], T.H. Lin[84], B.E. Lindquist[148], A.E. Lionti[51], E. Lipeles[122],
A. Lipniacka[15], M. Lisovyi[59b], T.M. Liss[165], A. Lister[167], A.M. Litke[137], B. Liu[151,ad], D. Liu[151],
H. Liu[90], H. Liu[27], J. Liu[86], J.B. Liu[35b], K. Liu[86], L. Liu[165], M. Liu[47], M. Liu[35b], Y.L. Liu[35b], Y. Liu[35b],
M. Livan[121a,121b], A. Lleres[57], J. Llorente Merino[35a], S.L. Lloyd[77], F. Lo Sterzo[151], E. Lobodzinska[44],
P. Loch[7], W.S. Lockman[137], F.K. Loebinger[85], A.E. Loevschall-Jensen[38], K.M. Loew[25], A. Loginov[175],
T. Lohse[17], K. Lohwasser[44], M. Lokajicek[127], B.A. Long[24], J.D. Long[165], R.E. Long[73], L. Longo[74a,74b],
K.A. Looper[111], L. Lopes[126a], D. Lopez Mateos[58], B. Lopez Paredes[139], I. Lopez Paz[13],
A. Lopez Solis[81], J. Lorenz[100], N. Lorenzo Martinez[62], M. Losada[21], P.J. Lösel[100], X. Lou[35a],
A. Lounis[117], J. Love[6], P.A. Love[73], H. Lu[61a], N. Lu[90], H.J. Lubatti[138], C. Luci[132a,132b], A. Lucotte[57],
C. Luedtke[50], F. Luehring[62], W. Lukas[63], L. Luminari[132a], O. Lundberg[146a,146b], B. Lund-Jensen[147],
P.M. Luzi[81], D. Lynn[27], R. Lysak[127], E. Lytken[82], V. Lyubushkin[66], H. Ma[27], L.L. Ma[35d], Y. Ma[35d],
G. Maccarrone[49], A. Macchiolo[101], C.M. Macdonald[139], B. Maček[76], J. Machado Miguens[122,126b],
D. Madaffari[86], R. Madar[36], H.J. Maddocks[164], W.F. Mader[46], A. Madsen[44], J. Maeda[68], S. Maeland[15],
T. Maeno[27], A. Maevskiy[99], E. Magradze[56], J. Mahlstedt[107], C. Maiani[117], C. Maidantchik[26a],
A.A. Maier[101], T. Maier[100], A. Maio[126a,126b,126d], S. Majewski[116], Y. Makida[67], N. Makovec[117],
B. Malaescu[81], Pa. Malecki[41], V.P. Maleev[123], F. Malek[57], U. Mallik[64], D. Malon[6], C. Malone[143],
S. Maltezos[10], S. Malyukov[32], J. Mamuzic[166], G. Mancini[49], B. Mandelli[32], L. Mandelli[92a], I. Mandić[76],
J. Maneira[126a,126b], L. Manhaes de Andrade Filho[26b], J. Manjarres Ramos[159b], A. Mann[100],
A. Manousos[32], B. Mansoulie[136], J.D. Mansour[35a], R. Mantifel[88], M. Mantoani[56], S. Manzoni[92a,92b],
L. Mapelli[32], G. Marceca[29], L. March[51], G. Marchiori[81], M. Marcisovsky[127], M. Marjanovic[14],
D.E. Marley[90], F. Marroquim[26a], S.P. Marsden[85], Z. Marshall[16], S. Marti-Garcia[166], B. Martin[91],
T.A. Martin[169], V.J. Martin[48], B. Martin dit Latour[15], M. Martinez[13,r], V.I. Martinez Outschoorn[165],
S. Martin-Haugh[131], V.S. Martoiu[28b], A.C. Martyniuk[79], M. Marx[138], A. Marzin[32], L. Masetti[84],
T. Mashimo[155], R. Mashinistov[96], J. Masik[85], A.L. Maslennikov[109,c], I. Massa[22a,22b], L. Massa[22a,22b],
P. Mastrandrea[5], A. Mastroberardino[39a,39b], T. Masubuchi[155], P. Mättig[174], J. Mattmann[84], J. Maurer[28b],
S.J. Maxfield[75], D.A. Maximov[109,c], R. Mazini[151], S.M. Mazza[92a,92b], N.C. Mc Fadden[105],
G. Mc Goldrick[158], S.P. Mc Kee[90], A. McCarn[90], R.L. McCarthy[148], T.G. McCarthy[101],
L.I. McClymont[79], E.F. McDonald[89], J.A. Mcfayden[79], G. Mchedlidze[56], S.J. McMahon[131],
R.A. McPherson[168,l], M. Medinnis[44], S. Meehan[138], S. Mehlhase[100], A. Mehta[75], K. Meier[59a],
C. Meineck[100], B. Meirose[43], D. Melini[166], B.R. Mellado Garcia[145c], M. Melo[144a], F. Meloni[18],
A. Mengarelli[22a,22b], S. Menke[101], E. Meoni[161], S. Mergelmeyer[17], P. Mermod[51], L. Merola[104a,104b],
C. Meroni[92a], F.S. Merritt[33], A. Messina[132a,132b], J. Metcalfe[6], A.S. Mete[162], C. Meyer[84], C. Meyer[122],
J-P. Meyer[136], J. Meyer[107], H. Meyer Zu Theenhausen[59a], F. Miano[149], R.P. Middleton[131],
S. Miglioranzi[52a,52b], L. Mijović[23], G. Mikenberg[171], M. Mikestikova[127], M. Mikuž[76], M. Milesi[89],
A. Milic[63], D.W. Miller[33], C. Mills[48], A. Milov[171], D.A. Milstead[146a,146b], A.A. Minaenko[130],
Y. Minami[155], I.A. Minashvili[66], A.I. Mincer[110], B. Mindur[40a], M. Mineev[66], Y. Ming[172], L.M. Mir[13],
K.P. Mistry[122], T. Mitani[170], J. Mitrevski[100], V.A. Mitsou[166], A. Miucci[51], P.S. Miyagawa[139],

J.U. Mjörnmark[82], T. Moa[146a,146b], K. Mochizuki[95], S. Mohapatra[37], S. Molander[146a,146b],
R. Moles-Valls[23], R. Monden[69], M.C. Mondragon[91], K. Mönig[44], J. Monk[38], E. Monnier[86],
A. Montalbano[148], J. Montejo Berlingen[32], F. Monticelli[72], S. Monzani[92a,92b], R.W. Moore[3],
N. Morange[117], D. Moreno[21], M. Moreno Llácer[56], P. Morettini[52a], D. Mori[142], T. Mori[155], M. Morii[58],
M. Morinaga[155], V. Morisbak[119], S. Moritz[84], A.K. Morley[150], G. Mornacchi[32], J.D. Morris[77],
S.S. Mortensen[38], L. Morvaj[148], M. Mosidze[53b], J. Moss[143], K. Motohashi[157], R. Mount[143],
E. Mountricha[27], S.V. Mouraviev[96,*], E.J.W. Moyse[87], S. Muanza[86], R.D. Mudd[19], F. Mueller[101],
J. Mueller[125], R.S.P. Mueller[100], T. Mueller[30], D. Muenstermann[73], P. Mullen[55], G.A. Mullier[18],
F.J. Munoz Sanchez[85], J.A. Murillo Quijada[19], W.J. Murray[169,131], H. Musheghyan[56], M. Muškinja[76],
A.G. Myagkov[130,ae], M. Myska[128], B.P. Nachman[143], O. Nackenhorst[51], K. Nagai[120], R. Nagai[67,z],
K. Nagano[67], Y. Nagasaka[60], K. Nagata[160], M. Nagel[50], E. Nagy[86], A.M. Nairz[32], Y. Nakahama[32],
K. Nakamura[67], T. Nakamura[155], I. Nakano[112], H. Namasivayam[43], R.F. Naranjo Garcia[44],
R. Narayan[11], D.I. Narrias Villar[59a], I. Naryshkin[123], T. Naumann[44], G. Navarro[21], R. Nayyar[7],
H.A. Neal[90], P.Yu. Nechaeva[96], T.J. Neep[85], P.D. Nef[143], A. Negri[121a,121b], M. Negrini[22a],
S. Nektarijevic[106], C. Nellist[117], A. Nelson[162], S. Nemecek[127], P. Nemethy[110], A.A. Nepomuceno[26a],
M. Nessi[32,af], M.S. Neubauer[165], M. Neumann[174], R.M. Neves[110], P. Nevski[27], P.R. Newman[19],
D.H. Nguyen[6], T. Nguyen Manh[95], R.B. Nickerson[120], R. Nicolaidou[136], J. Nielsen[137], A. Nikiforov[17],
V. Nikolaenko[130,ae], I. Nikolic-Audit[81], K. Nikolopoulos[19], J.K. Nilsen[119], P. Nilsson[27], Y. Ninomiya[155],
A. Nisati[132a], R. Nisius[101], T. Nobe[155], M. Nomachi[118], I. Nomidis[31], T. Nooney[77], S. Norberg[113],
M. Nordberg[32], N. Norjoharuddeen[120], O. Novgorodova[46], S. Nowak[101], M. Nozaki[67], L. Nozka[115],
K. Ntekas[10], E. Nurse[79], F. Nuti[89], F. O'grady[7], D.C. O'Neil[142], A.A. O'Rourke[44], V. O'Shea[55],
F.G. Oakham[31,d], H. Oberlack[101], T. Obermann[23], J. Ocariz[81], A. Ochi[68], I. Ochoa[37],
J.P. Ochoa-Ricoux[34a], S. Oda[71], S. Odaka[67], H. Ogren[62], A. Oh[85], S.H. Oh[47], C.C. Ohm[16],
H. Ohman[164], H. Oide[32], H. Okawa[160], Y. Okumura[33], T. Okuyama[67], A. Olariu[28b],
L.F. Oleiro Seabra[126a], S.A. Olivares Pino[48], D. Oliveira Damazio[27], A. Olszewski[41], J. Olszowska[41],
A. Onofre[126a,126e], K. Onogi[103], P.U.E. Onyisi[11,v], M.J. Oreglia[33], Y. Oren[153], D. Orestano[134a,134b],
N. Orlando[61b], R.S. Orr[158], B. Osculati[52a,52b], R. Ospanov[85], G. Otero y Garzon[29], H. Otono[71],
M. Ouchrif[135d], F. Ould-Saada[119], A. Ouraou[136], K.P. Oussoren[107], Q. Ouyang[35a], M. Owen[55],
R.E. Owen[19], V.E. Ozcan[20a], N. Ozturk[8], K. Pachal[142], A. Pacheco Pages[13], L. Pacheco Rodriguez[136],
C. Padilla Aranda[13], M. Pagáčová[50], S. Pagan Griso[16], F. Paige[27], P. Pais[87], K. Pajchel[119],
G. Palacino[159b], S. Palestini[32], M. Palka[40b], D. Pallin[36], A. Palma[126a,126b], E.St. Panagiotopoulou[10],
C.E. Pandini[81], J.G. Panduro Vazquez[78], P. Pani[146a,146b], S. Panitkin[27], D. Pantea[28b], L. Paolozzi[51],
Th.D. Papadopoulou[10], K. Papageorgiou[154], A. Paramonov[6], D. Paredes Hernandez[175], A.J. Parker[73],
M.A. Parker[30], K.A. Parker[139], F. Parodi[52a,52b], J.A. Parsons[37], U. Parzefall[50], V.R. Pascuzzi[158],
E. Pasqualucci[132a], S. Passaggio[52a], Fr. Pastore[78], G. Pásztor[31,ag], S. Pataraia[174], J.R. Pater[85], T. Pauly[32],
J. Pearce[168], B. Pearson[113], L.E. Pedersen[38], M. Pedersen[119], S. Pedraza Lopez[166], R. Pedro[126a,126b],
S.V. Peleganchuk[109,c], D. Pelikan[164], O. Penc[127], C. Peng[35a], H. Peng[35b], J. Penwell[62], B.S. Peralva[26b],
M.M. Perego[136], D.V. Perepelitsa[27], E. Perez Codina[159a], L. Perini[92a,92b], H. Pernegger[32],
S. Perrella[104a,104b], R. Peschke[44], V.D. Peshekhonov[66], K. Peters[44], R.F.Y. Peters[85], B.A. Petersen[32],
T.C. Petersen[38], E. Petit[57], A. Petridis[1], C. Petridou[154], P. Petroff[117], E. Petrolo[132a], M. Petrov[120],
F. Petrucci[134a,134b], N.E. Pettersson[87], A. Peyaud[136], R. Pezoa[34b], P.W. Phillips[131], G. Piacquadio[143],
E. Pianori[169], A. Picazio[87], E. Piccaro[77], M. Piccinini[22a,22b], M.A. Pickering[120], R. Piegaia[29],
J.E. Pilcher[33], A.D. Pilkington[85], A.W.J. Pin[85], M. Pinamonti[163a,163c,ah], J.L. Pinfold[3], A. Pingel[38],
S. Pires[81], H. Pirumov[44], M. Pitt[171], L. Plazak[144a], M.-A. Pleier[27], V. Pleskot[84], E. Plotnikova[66],
P. Plucinski[91], D. Pluth[65], R. Poettgen[146a,146b], L. Poggioli[117], D. Pohl[23], G. Polesello[121a], A. Poley[44],
A. Policicchio[39a,39b], R. Polifka[158], A. Polini[22a], C.S. Pollard[55], V. Polychronakos[27], K. Pommès[32],
L. Pontecorvo[132a], B.G. Pope[91], G.A. Popeneciu[28c], D.S. Popovic[14], A. Poppleton[32], S. Pospisil[128],

K. Potamianos[16], I.N. Potrap[66], C.J. Potter[30], C.T. Potter[116], G. Poulard[32], J. Poveda[32],
V. Pozdnyakov[66], M.E. Pozo Astigarraga[32], P. Pralavorio[86], A. Pranko[16], S. Prell[65], D. Price[85],
L.E. Price[6], M. Primavera[74a], S. Prince[88], M. Proissl[48], K. Prokofiev[61c], F. Prokoshin[34b],
S. Protopopescu[27], J. Proudfoot[6], M. Przybycien[40a], D. Puddu[134a,134b], M. Purohit[27,ai], P. Puzo[117],
J. Qian[90], G. Qin[55], Y. Qin[85], A. Quadt[56], W.B. Quayle[163a,163b], M. Queitsch-Maitland[85], D. Quilty[55],
S. Raddum[119], V. Radeka[27], V. Radescu[59b], S.K. Radhakrishnan[148], P. Radloff[116], P. Rados[89],
F. Ragusa[92a,92b], G. Rahal[177], J.A. Raine[85], S. Rajagopalan[27], M. Rammensee[32], C. Rangel-Smith[164],
M.G. Ratti[92a,92b], F. Rauscher[100], S. Rave[84], T. Ravenscroft[55], I. Ravinovich[171], M. Raymond[32],
A.L. Read[119], N.P. Readioff[75], M. Reale[74a,74b], D.M. Rebuzzi[121a,121b], A. Redelbach[173], G. Redlinger[27],
R. Reece[137], K. Reeves[43], L. Rehnisch[17], J. Reichert[122], H. Reisin[29], C. Rembser[32], H. Ren[35a],
M. Rescigno[132a], S. Resconi[92a], O.L. Rezanova[109,c], P. Reznicek[129], R. Rezvani[95], R. Richter[101],
S. Richter[79], E. Richter-Was[40b], O. Ricken[23], M. Ridel[81], P. Rieck[17], C.J. Riegel[174], J. Rieger[56],
O. Rifki[113], M. Rijssenbeek[148], A. Rimoldi[121a,121b], M. Rimoldi[18], L. Rinaldi[22a], B. Ristić[51], E. Ritsch[32],
I. Riu[13], F. Rizatdinova[114], E. Rizvi[77], C. Rizzi[13], S.H. Robertson[88,l], A. Robichaud-Veronneau[88],
D. Robinson[30], J.E.M. Robinson[44], A. Robson[55], C. Roda[124a,124b], Y. Rodina[86], A. Rodriguez Perez[13],
D. Rodriguez Rodriguez[166], S. Roe[32], C.S. Rogan[58], O. Røhne[119], A. Romaniouk[98], M. Romano[22a,22b],
S.M. Romano Saez[36], E. Romero Adam[166], N. Rompotis[138], M. Ronzani[50], L. Roos[81], E. Ros[166],
S. Rosati[132a], K. Rosbach[50], P. Rose[137], O. Rosenthal[141], N.-A. Rosien[56], V. Rossetti[146a,146b],
E. Rossi[104a,104b], L.P. Rossi[52a], J.H.N. Rosten[30], R. Rosten[138], M. Rotaru[28b], I. Roth[171], J. Rothberg[138],
D. Rousseau[117], C.R. Royon[136], A. Rozanov[86], Y. Rozen[152], X. Ruan[145c], F. Rubbo[143],
M.S. Rudolph[158], F. Rühr[50], A. Ruiz-Martinez[31], Z. Rurikova[50], N.A. Rusakovich[66], A. Ruschke[100],
H.L. Russell[138], J.P. Rutherfoord[7], N. Ruthmann[32], Y.F. Ryabov[123], M. Rybar[165], G. Rybkin[117], S. Ryu[6],
A. Ryzhov[130], G.F. Rzehorz[56], A.F. Saavedra[150], G. Sabato[107], S. Sacerdoti[29], H.F-W. Sadrozinski[137],
R. Sadykov[66], F. Safai Tehrani[132a], P. Saha[108], M. Sahinsoy[59a], M. Saimpert[136], T. Saito[155],
H. Sakamoto[155], Y. Sakurai[170], G. Salamanna[134a,134b], A. Salamon[133a,133b], J.E. Salazar Loyola[34b],
D. Salek[107], P.H. Sales De Bruin[138], D. Salihagic[101], A. Salnikov[143], J. Salt[166], D. Salvatore[39a,39b],
F. Salvatore[149], A. Salvucci[61a], A. Salzburger[32], D. Sammel[50], D. Sampsonidis[154], A. Sanchez[104a,104b],
J. Sánchez[166], V. Sanchez Martinez[166], H. Sandaker[119], R.L. Sandbach[77], H.G. Sander[84],
M. Sandhoff[174], C. Sandoval[21], R. Sandstroem[101], D.P.C. Sankey[131], M. Sannino[52a,52b], A. Sansoni[49],
C. Santoni[36], R. Santonico[133a,133b], H. Santos[126a], I. Santoyo Castillo[149], K. Sapp[125], A. Sapronov[66],
J.G. Saraiva[126a,126d], B. Sarrazin[23], O. Sasaki[67], Y. Sasaki[155], K. Sato[160], G. Sauvage[5,*], E. Sauvan[5],
G. Savage[78], P. Savard[158,d], C. Sawyer[131], L. Sawyer[80,q], J. Saxon[33], C. Sbarra[22a], A. Sbrizzi[22a,22b],
T. Scanlon[79], D.A. Scannicchio[162], M. Scarcella[150], V. Scarfone[39a,39b], J. Schaarschmidt[171],
P. Schacht[101], B.M. Schachtner[100], D. Schaefer[32], R. Schaefer[44], J. Schaeffer[84], S. Schaepe[23],
S. Schaetzel[59b], U. Schäfer[84], A.C. Schaffer[117], D. Schaile[100], R.D. Schamberger[148], V. Scharf[59a],
V.A. Schegelsky[123], D. Scheirich[129], M. Schernau[162], C. Schiavi[52a,52b], S. Schier[137], C. Schillo[50],
M. Schioppa[39a,39b], S. Schlenker[32], K.R. Schmidt-Sommerfeld[101], K. Schmieden[32], C. Schmitt[84],
S. Schmitt[44], S. Schmitz[84], B. Schneider[159a], U. Schnoor[50], L. Schoeffel[136], A. Schoening[59b],
B.D. Schoenrock[91], E. Schopf[23], M. Schott[84], J. Schovancova[8], S. Schramm[51], M. Schreyer[173],
N. Schuh[84], A. Schulte[84], M.J. Schultens[23], H.-C. Schultz-Coulon[59a], H. Schulz[17], M. Schumacher[50],
B.A. Schumm[137], Ph. Schune[136], A. Schwartzman[143], T.A. Schwarz[90], Ph. Schwegler[101],
H. Schweiger[85], Ph. Schwemling[136], R. Schwienhorst[91], J. Schwindling[136], T. Schwindt[23], G. Sciolla[25],
F. Scuri[124a,124b], F. Scutti[89], J. Searcy[90], P. Seema[23], S.C. Seidel[105], A. Seiden[137], F. Seifert[128],
J.M. Seixas[26a], G. Sekhniaidze[104a], K. Sekhon[90], S.J. Sekula[42], D.M. Seliverstov[123,*],
N. Semprini-Cesari[22a,22b], C. Serfon[119], L. Serin[117], L. Serkin[163a,163b], M. Sessa[134a,134b], R. Seuster[168],
H. Severini[113], T. Sfiligoj[76], F. Sforza[32], A. Sfyrla[51], E. Shabalina[56], N.W. Shaikh[146a,146b], L.Y. Shan[35a],
R. Shang[165], J.T. Shank[24], M. Shapiro[16], P.B. Shatalov[97], K. Shaw[163a,163b], S.M. Shaw[85],

A. Shcherbakova[146a,146b], C.Y. Shehu[149], P. Sherwood[79], L. Shi[151,aj], S. Shimizu[68], C.O. Shimmin[162],
M. Shimojima[102], M. Shiyakova[66,ak], A. Shmeleva[96], D. Shoaleh Saadi[95], M.J. Shochet[33],
S. Shojaii[92a,92b], S. Shrestha[111], E. Shulga[98], M.A. Shupe[7], P. Sicho[127], A.M. Sickles[165], P.E. Sidebo[147],
O. Sidiropoulou[173], D. Sidorov[114], A. Sidoti[22a,22b], F. Siegert[46], Dj. Sijacki[14], J. Silva[126a,126d],
S.B. Silverstein[146a], V. Simak[128], O. Simard[5], Lj. Simic[14], S. Simion[117], E. Simioni[84], B. Simmons[79],
D. Simon[36], M. Simon[84], P. Sinervo[158], N.B. Sinev[116], M. Sioli[22a,22b], G. Siragusa[173],
S.Yu. Sivoklokov[99], J. Sjölin[146a,146b], M.B. Skinner[73], H.P. Skottowe[58], P. Skubic[113], M. Slater[19],
T. Slavicek[128], M. Slawinska[107], K. Sliwa[161], R. Slovak[129], V. Smakhtin[171], B.H. Smart[5], L. Smestad[15],
J. Smiesko[144a], S.Yu. Smirnov[98], Y. Smirnov[98], L.N. Smirnova[99,al], O. Smirnova[82], M.N.K. Smith[37],
R.W. Smith[37], M. Smizanska[73], K. Smolek[128], A.A. Snesarev[96], S. Snyder[27], R. Sobie[168,l], F. Socher[46],
A. Soffer[153], D.A. Soh[151], G. Sokhrannyi[76], C.A. Solans Sanchez[32], M. Solar[128], E.Yu. Soldatov[98],
U. Soldevila[166], A.A. Solodkov[130], A. Soloshenko[66], O.V. Solovyanov[130], V. Solovyev[123], P. Sommer[50],
H. Son[161], H.Y. Song[35b,am], A. Sood[16], A. Sopczak[128], V. Sopko[128], V. Sorin[13], D. Sosa[59b],
C.L. Sotiropoulou[124a,124b], R. Soualah[163a,163c], A.M. Soukharev[109,c], D. South[44], B.C. Sowden[78],
S. Spagnolo[74a,74b], M. Spalla[124a,124b], M. Spangenberg[169], F. Spanò[78], D. Sperlich[17], F. Spettel[101],
R. Spighi[22a], G. Spigo[32], L.A. Spiller[89], M. Spousta[129], R.D. St. Denis[55,*], A. Stabile[92a], R. Stamen[59a],
S. Stamm[17], E. Stanecka[41], R.W. Stanek[6], C. Stanescu[134a], M. Stanescu-Bellu[44], M.M. Stanitzki[44],
S. Stapnes[119], E.A. Starchenko[130], G.H. Stark[33], J. Stark[57], P. Staroba[127], P. Starovoitov[59a], S. Stärz[32],
R. Staszewski[41], P. Steinberg[27], B. Stelzer[142], H.J. Stelzer[32], O. Stelzer-Chilton[159a], H. Stenzel[54],
G.A. Stewart[55], J.A. Stillings[23], M.C. Stockton[88], M. Stoebe[88], G. Stoicea[28b], P. Stolte[56], S. Stonjek[101],
A.R. Stradling[8], A. Straessner[46], M.E. Stramaglia[18], J. Strandberg[147], S. Strandberg[146a,146b],
A. Strandlie[119], M. Strauss[113], P. Strizenec[144b], R. Ströhmer[173], D.M. Strom[116], R. Stroynowski[42],
A. Strubig[106], S.A. Stucci[18], B. Stugu[15], N.A. Styles[44], D. Su[143], J. Su[125], S. Suchek[59a], Y. Sugaya[118],
M. Suk[128], V.V. Sulin[96], S. Sultansoy[4c], T. Sumida[69], S. Sun[58], X. Sun[35a], J.E. Sundermann[50],
K. Suruliz[149], G. Susinno[39a,39b], M.R. Sutton[149], S. Suzuki[67], M. Svatos[127], M. Swiatlowski[33],
I. Sykora[144a], T. Sykora[129], D. Ta[50], C. Taccini[134a,134b], K. Tackmann[44], J. Taenzer[158], A. Taffard[162],
R. Tafirout[159a], N. Taiblum[153], H. Takai[27], R. Takashima[70], T. Takeshita[140], Y. Takubo[67], M. Talby[86],
A.A. Talyshev[109,c], K.G. Tan[89], J. Tanaka[155], R. Tanaka[117], S. Tanaka[67], B.B. Tannenwald[111],
S. Tapia Araya[34b], S. Tapprogge[84], S. Tarem[152], G.F. Tartarelli[92a], P. Tas[129], M. Tasevsky[127],
T. Tashiro[69], E. Tassi[39a,39b], A. Tavares Delgado[126a,126b], Y. Tayalati[135d], A.C. Taylor[105], G.N. Taylor[89],
P.T.E. Taylor[89], W. Taylor[159b], F.A. Teischinger[32], P. Teixeira-Dias[78], K.K. Temming[50], D. Temple[142],
H. Ten Kate[32], P.K. Teng[151], J.J. Teoh[118], F. Tepel[174], S. Terada[67], K. Terashi[155], J. Terron[83], S. Terzo[101],
M. Testa[49], R.J. Teuscher[158,l], T. Theveneaux-Pelzer[86], J.P. Thomas[19], J. Thomas-Wilsker[78],
E.N. Thompson[37], P.D. Thompson[19], A.S. Thompson[55], L.A. Thomsen[175], E. Thomson[122],
M. Thomson[30], M.J. Tibbetts[16], R.E. Ticse Torres[86], V.O. Tikhomirov[96,an], Yu.A. Tikhonov[109,c],
S. Timoshenko[98], P. Tipton[175], S. Tisserant[86], K. Todome[157], T. Todorov[5,*], S. Todorova-Nova[129],
J. Tojo[71], S. Tokár[144a], K. Tokushuku[67], E. Tolley[58], L. Tomlinson[85], M. Tomoto[103], L. Tompkins[143,ao],
K. Toms[105], B. Tong[58], E. Torrence[116], H. Torres[142], E. Torró Pastor[138], J. Toth[86,ap], F. Touchard[86],
D.R. Tovey[139], T. Trefzger[173], A. Tricoli[27], I.M. Trigger[159a], S. Trincaz-Duvoid[81], M.F. Tripiana[13],
W. Trischuk[158], B. Trocmé[57], A. Trofymov[44], C. Troncon[92a], M. Trottier-McDonald[16], M. Trovatelli[168],
L. Truong[163a,163c], M. Trzebinski[41], A. Trzupek[41], J.C-L. Tseng[120], P.V. Tsiareshka[93], G. Tsipolitis[10],
N. Tsirintanis[9], S. Tsiskaridze[13], V. Tsiskaridze[50], E.G. Tskhadadze[53a], K.M. Tsui[61a], I.I. Tsukerman[97],
V. Tsulaia[16], S. Tsuno[67], D. Tsybychev[148], A. Tudorache[28b], V. Tudorache[28b], A.N. Tuna[58],
S.A. Tupputi[22a,22b], S. Turchikhin[99,al], D. Turecek[128], D. Turgeman[171], R. Turra[92a,92b], A.J. Turvey[42],
P.M. Tuts[37], M. Tyndel[131], G. Ucchielli[22a,22b], I. Ueda[155], M. Ughetto[146a,146b], F. Ukegawa[160],
G. Unal[32], A. Undrus[27], G. Unel[162], F.C. Ungaro[89], Y. Unno[67], C. Unverdorben[100], J. Urban[144b],
P. Urquijo[89], P. Urrejola[84], G. Usai[8], A. Usanova[63], L. Vacavant[86], V. Vacek[128], B. Vachon[88],

C. Valderanis[100], E. Valdes Santurio[146a,146b], N. Valencic[107], S. Valentinetti[22a,22b], A. Valero[166],
L. Valery[13], S. Valkar[129], S. Vallecorsa[51], J.A. Valls Ferrer[166], W. Van Den Wollenberg[107],
P.C. Van Der Deijl[107], R. van der Geer[107], H. van der Graaf[107], N. van Eldik[152], P. van Gemmeren[6],
J. Van Nieuwkoop[142], I. van Vulpen[107], M.C. van Woerden[32], M. Vanadia[132a,132b], W. Vandelli[32],
R. Vanguri[122], A. Vaniachine[130], P. Vankov[107], G. Vardanyan[176], R. Vari[132a], E.W. Varnes[7], T. Varol[42],
D. Varouchas[81], A. Vartapetian[8], K.E. Varvell[150], J.G. Vasquez[175], F. Vazeille[36],
T. Vazquez Schroeder[88], J. Veatch[56], L.M. Veloce[158], F. Veloso[126a,126c], S. Veneziano[132a],
A. Ventura[74a,74b], M. Venturi[168], N. Venturi[158], A. Venturini[25], V. Vercesi[121a], M. Verducci[132a,132b],
W. Verkerke[107], J.C. Vermeulen[107], A. Vest[46,aq], M.C. Vetterli[142,d], O. Viazlo[82], I. Vichou[165],
T. Vickey[139], O.E. Vickey Boeriu[139], G.H.A. Viehhauser[120], S. Viel[16], L. Vigani[120], R. Vigne[63],
M. Villa[22a,22b], M. Villaplana Perez[92a,92b], E. Vilucchi[49], M.G. Vincter[31], V.B. Vinogradov[66],
C. Vittori[22a,22b], I. Vivarelli[149], S. Vlachos[10], M. Vlasak[128], M. Vogel[174], P. Vokac[128], G. Volpi[124a,124b],
M. Volpi[89], H. von der Schmitt[101], E. von Toerne[23], V. Vorobel[129], K. Vorobev[98], M. Vos[166], R. Voss[32],
J.H. Vossebeld[75], N. Vranjes[14], M. Vranjes Milosavljevic[14], V. Vrba[127], M. Vreeswijk[107],
R. Vuillermet[32], I. Vukotic[33], Z. Vykydal[128], P. Wagner[23], W. Wagner[174], H. Wahlberg[72],
S. Wahrmund[46], J. Wakabayashi[103], J. Walder[73], R. Walker[100], W. Walkowiak[141], V. Wallangen[146a,146b],
C. Wang[35c], C. Wang[35d,86], F. Wang[172], H. Wang[16], H. Wang[42], J. Wang[44], J. Wang[150], K. Wang[88],
R. Wang[6], S.M. Wang[151], T. Wang[23], T. Wang[37], W. Wang[35b], X. Wang[175], C. Wanotayaroj[116],
A. Warburton[88], C.P. Ward[30], D.R. Wardrope[79], A. Washbrook[48], P.M. Watkins[19], A.T. Watson[19],
M.F. Watson[19], G. Watts[138], S. Watts[85], B.M. Waugh[79], S. Webb[84], M.S. Weber[18], S.W. Weber[173],
J.S. Webster[6], A.R. Weidberg[120], B. Weinert[62], J. Weingarten[56], C. Weiser[50], H. Weits[107], P.S. Wells[32],
T. Wenaus[27], T. Wengler[32], S. Wenig[32], N. Wermes[23], M. Werner[50], M.D. Werner[65], P. Werner[32],
M. Wessels[59a], J. Wetter[161], K. Whalen[116], N.L. Whallon[138], A.M. Wharton[73], A. White[8], M.J. White[1],
R. White[34b], D. Whiteson[162], F.J. Wickens[131], W. Wiedenmann[172], M. Wielers[131], P. Wienemann[23],
C. Wiglesworth[38], L.A.M. Wiik-Fuchs[23], A. Wildauer[101], F. Wilk[85], H.G. Wilkens[32], H.H. Williams[122],
S. Williams[107], C. Willis[91], S. Willocq[87], J.A. Wilson[19], I. Wingerter-Seez[5], F. Winklmeier[116],
O.J. Winston[149], B.T. Winter[23], M. Wittgen[143], J. Wittkowski[100], M.W. Wolter[41], H. Wolters[126a,126c],
S.D. Worm[131], B.K. Wosiek[41], J. Wotschack[32], M.J. Woudstra[85], K.W. Wozniak[41], M. Wu[57], M. Wu[33],
S.L. Wu[172], X. Wu[51], Y. Wu[90], T.R. Wyatt[85], B.M. Wynne[48], S. Xella[38], D. Xu[35a], L. Xu[27],
B. Yabsley[150], S. Yacoob[145a], R. Yakabe[68], D. Yamaguchi[157], Y. Yamaguchi[118], A. Yamamoto[67],
S. Yamamoto[155], T. Yamanaka[155], K. Yamauchi[103], Y. Yamazaki[68], Z. Yan[24], H. Yang[35e], H. Yang[172],
Y. Yang[151], Z. Yang[15], W-M. Yao[16], Y.C. Yap[81], Y. Yasu[67], E. Yatsenko[5], K.H. Yau Wong[23], J. Ye[42],
S. Ye[27], I. Yeletskikh[66], A.L. Yen[58], E. Yildirim[84], K. Yorita[170], R. Yoshida[6], K. Yoshihara[122],
C. Young[143], C.J.S. Young[32], S. Youssef[24], D.R. Yu[16], J. Yu[8], J.M. Yu[90], J. Yu[65], L. Yuan[68],
S.P.Y. Yuen[23], I. Yusuff[30,ar], B. Zabinski[41], R. Zaidan[35d], A.M. Zaitsev[130,ae], N. Zakharchuk[44],
J. Zalieckas[15], A. Zaman[148], S. Zambito[58], L. Zanello[132a,132b], D. Zanzi[89], C. Zeitnitz[174], M. Zeman[128],
A. Zemla[40a], J.C. Zeng[165], Q. Zeng[143], K. Zengel[25], O. Zenin[130], T. Ženiš[144a], D. Zerwas[117],
D. Zhang[90], F. Zhang[172], G. Zhang[35b,am], H. Zhang[35c], J. Zhang[6], L. Zhang[50], R. Zhang[23],
R. Zhang[35b,as], X. Zhang[35d], Z. Zhang[117], X. Zhao[42], Y. Zhao[35d], Z. Zhao[35b], A. Zhemchugov[66],
J. Zhong[120], B. Zhou[90], C. Zhou[47], L. Zhou[37], L. Zhou[42], M. Zhou[148], N. Zhou[35f], C.G. Zhu[35d],
H. Zhu[35a], J. Zhu[90], Y. Zhu[35b], X. Zhuang[35a], K. Zhukov[96], A. Zibell[173], D. Zieminska[62], N.I. Zimine[66],
C. Zimmermann[84], S. Zimmermann[50], Z. Zinonos[56], M. Zinser[84], M. Ziolkowski[141], L. Živković[14],
G. Zobernig[172], A. Zoccoli[22a,22b], M. zur Nedden[17], L. Zwalinski[32].

[1] Department of Physics, University of Adelaide, Adelaide, Australia
[2] Physics Department, SUNY Albany, Albany NY, United States of America
[3] Department of Physics, University of Alberta, Edmonton AB, Canada

[4] [a] Department of Physics, Ankara University, Ankara; [b] Istanbul Aydin University, Istanbul; [c] Division of Physics, TOBB University of Economics and Technology, Ankara, Turkey
[5] LAPP, CNRS/IN2P3 and Université Savoie Mont Blanc, Annecy-le-Vieux, France
[6] High Energy Physics Division, Argonne National Laboratory, Argonne IL, United States of America
[7] Department of Physics, University of Arizona, Tucson AZ, United States of America
[8] Department of Physics, The University of Texas at Arlington, Arlington TX, United States of America
[9] Physics Department, University of Athens, Athens, Greece
[10] Physics Department, National Technical University of Athens, Zografou, Greece
[11] Department of Physics, The University of Texas at Austin, Austin TX, United States of America
[12] Institute of Physics, Azerbaijan Academy of Sciences, Baku, Azerbaijan
[13] Institut de Física d'Altes Energies (IFAE), The Barcelona Institute of Science and Technology, Barcelona, Spain, Spain
[14] Institute of Physics, University of Belgrade, Belgrade, Serbia
[15] Department for Physics and Technology, University of Bergen, Bergen, Norway
[16] Physics Division, Lawrence Berkeley National Laboratory and University of California, Berkeley CA, United States of America
[17] Department of Physics, Humboldt University, Berlin, Germany
[18] Albert Einstein Center for Fundamental Physics and Laboratory for High Energy Physics, University of Bern, Bern, Switzerland
[19] School of Physics and Astronomy, University of Birmingham, Birmingham, United Kingdom
[20] [a] Department of Physics, Bogazici University, Istanbul; [b] Department of Physics Engineering, Gaziantep University, Gaziantep; [d] Istanbul Bilgi University, Faculty of Engineering and Natural Sciences, Istanbul,Turkey; [e] Bahcesehir University, Faculty of Engineering and Natural Sciences, Istanbul, Turkey, Turkey
[21] Centro de Investigaciones, Universidad Antonio Narino, Bogota, Colombia
[22] [a] INFN Sezione di Bologna; [b] Dipartimento di Fisica e Astronomia, Università di Bologna, Bologna, Italy
[23] Physikalisches Institut, University of Bonn, Bonn, Germany
[24] Department of Physics, Boston University, Boston MA, United States of America
[25] Department of Physics, Brandeis University, Waltham MA, United States of America
[26] [a] Universidade Federal do Rio De Janeiro COPPE/EE/IF, Rio de Janeiro; [b] Electrical Circuits Department, Federal University of Juiz de Fora (UFJF), Juiz de Fora; [c] Federal University of Sao Joao del Rei (UFSJ), Sao Joao del Rei; [d] Instituto de Fisica, Universidade de Sao Paulo, Sao Paulo, Brazil
[27] Physics Department, Brookhaven National Laboratory, Upton NY, United States of America
[28] [a] Transilvania University of Brasov, Brasov, Romania; [b] National Institute of Physics and Nuclear Engineering, Bucharest; [c] National Institute for Research and Development of Isotopic and Molecular Technologies, Physics Department, Cluj Napoca; [d] University Politehnica Bucharest, Bucharest; [e] West University in Timisoara, Timisoara, Romania
[29] Departamento de Física, Universidad de Buenos Aires, Buenos Aires, Argentina
[30] Cavendish Laboratory, University of Cambridge, Cambridge, United Kingdom
[31] Department of Physics, Carleton University, Ottawa ON, Canada
[32] CERN, Geneva, Switzerland
[33] Enrico Fermi Institute, University of Chicago, Chicago IL, United States of America
[34] [a] Departamento de Física, Pontificia Universidad Católica de Chile, Santiago; [b] Departamento de Física, Universidad Técnica Federico Santa María, Valparaíso, Chile
[35] [a] Institute of High Energy Physics, Chinese Academy of Sciences, Beijing; [b] Department of Modern Physics, University of Science and Technology of China, Anhui; [c] Department of Physics,

Nanjing University, Jiangsu; [d] School of Physics, Shandong University, Shandong; [e] Department of Physics and Astronomy, Shanghai Key Laboratory for Particle Physics and Cosmology, Shanghai Jiao Tong University, Shanghai; (also affiliated with PKU-CHEP); [f] Physics Department, Tsinghua University, Beijing 100084, China

36 Laboratoire de Physique Corpusculaire, Clermont Université and Université Blaise Pascal and CNRS/IN2P3, Clermont-Ferrand, France

37 Nevis Laboratory, Columbia University, Irvington NY, United States of America

38 Niels Bohr Institute, University of Copenhagen, Kobenhavn, Denmark

39 [a] INFN Gruppo Collegato di Cosenza, Laboratori Nazionali di Frascati; [b] Dipartimento di Fisica, Università della Calabria, Rende, Italy

40 [a] AGH University of Science and Technology, Faculty of Physics and Applied Computer Science, Krakow; [b] Marian Smoluchowski Institute of Physics, Jagiellonian University, Krakow, Poland

41 Institute of Nuclear Physics Polish Academy of Sciences, Krakow, Poland

42 Physics Department, Southern Methodist University, Dallas TX, United States of America

43 Physics Department, University of Texas at Dallas, Richardson TX, United States of America

44 DESY, Hamburg and Zeuthen, Germany

45 Institut für Experimentelle Physik IV, Technische Universität Dortmund, Dortmund, Germany

46 Institut für Kern- und Teilchenphysik, Technische Universität Dresden, Dresden, Germany

47 Department of Physics, Duke University, Durham NC, United States of America

48 SUPA - School of Physics and Astronomy, University of Edinburgh, Edinburgh, United Kingdom

49 INFN Laboratori Nazionali di Frascati, Frascati, Italy

50 Fakultät für Mathematik und Physik, Albert-Ludwigs-Universität, Freiburg, Germany

51 Section de Physique, Université de Genève, Geneva, Switzerland

52 [a] INFN Sezione di Genova; [b] Dipartimento di Fisica, Università di Genova, Genova, Italy

53 [a] E. Andronikashvili Institute of Physics, Iv. Javakhishvili Tbilisi State University, Tbilisi; [b] High Energy Physics Institute, Tbilisi State University, Tbilisi, Georgia

54 II Physikalisches Institut, Justus-Liebig-Universität Giessen, Giessen, Germany

55 SUPA - School of Physics and Astronomy, University of Glasgow, Glasgow, United Kingdom

56 II Physikalisches Institut, Georg-August-Universität, Göttingen, Germany

57 Laboratoire de Physique Subatomique et de Cosmologie, Université Grenoble-Alpes, CNRS/IN2P3, Grenoble, France

58 Laboratory for Particle Physics and Cosmology, Harvard University, Cambridge MA, United States of America

59 [a] Kirchhoff-Institut für Physik, Ruprecht-Karls-Universität Heidelberg, Heidelberg; [b] Physikalisches Institut, Ruprecht-Karls-Universität Heidelberg, Heidelberg; [c] ZITI Institut für technische Informatik, Ruprecht-Karls-Universität Heidelberg, Mannheim, Germany

60 Faculty of Applied Information Science, Hiroshima Institute of Technology, Hiroshima, Japan

61 [a] Department of Physics, The Chinese University of Hong Kong, Shatin, N.T., Hong Kong; [b] Department of Physics, The University of Hong Kong, Hong Kong; [c] Department of Physics, The Hong Kong University of Science and Technology, Clear Water Bay, Kowloon, Hong Kong, China

62 Department of Physics, Indiana University, Bloomington IN, United States of America

63 Institut für Astro- und Teilchenphysik, Leopold-Franzens-Universität, Innsbruck, Austria

64 University of Iowa, Iowa City IA, United States of America

65 Department of Physics and Astronomy, Iowa State University, Ames IA, United States of America

66 Joint Institute for Nuclear Research, JINR Dubna, Dubna, Russia

67 KEK, High Energy Accelerator Research Organization, Tsukuba, Japan

68 Graduate School of Science, Kobe University, Kobe, Japan

[69] Faculty of Science, Kyoto University, Kyoto, Japan
[70] Kyoto University of Education, Kyoto, Japan
[71] Department of Physics, Kyushu University, Fukuoka, Japan
[72] Instituto de Física La Plata, Universidad Nacional de La Plata and CONICET, La Plata, Argentina
[73] Physics Department, Lancaster University, Lancaster, United Kingdom
[74] [a] INFN Sezione di Lecce; [b] Dipartimento di Matematica e Fisica, Università del Salento, Lecce, Italy
[75] Oliver Lodge Laboratory, University of Liverpool, Liverpool, United Kingdom
[76] Department of Physics, Jožef Stefan Institute and University of Ljubljana, Ljubljana, Slovenia
[77] School of Physics and Astronomy, Queen Mary University of London, London, United Kingdom
[78] Department of Physics, Royal Holloway University of London, Surrey, United Kingdom
[79] Department of Physics and Astronomy, University College London, London, United Kingdom
[80] Louisiana Tech University, Ruston LA, United States of America
[81] Laboratoire de Physique Nucléaire et de Hautes Energies, UPMC and Université Paris-Diderot and CNRS/IN2P3, Paris, France
[82] Fysiska institutionen, Lunds universitet, Lund, Sweden
[83] Departamento de Fisica Teorica C-15, Universidad Autonoma de Madrid, Madrid, Spain
[84] Institut für Physik, Universität Mainz, Mainz, Germany
[85] School of Physics and Astronomy, University of Manchester, Manchester, United Kingdom
[86] CPPM, Aix-Marseille Université and CNRS/IN2P3, Marseille, France
[87] Department of Physics, University of Massachusetts, Amherst MA, United States of America
[88] Department of Physics, McGill University, Montreal QC, Canada
[89] School of Physics, University of Melbourne, Victoria, Australia
[90] Department of Physics, The University of Michigan, Ann Arbor MI, United States of America
[91] Department of Physics and Astronomy, Michigan State University, East Lansing MI, United States of America
[92] [a] INFN Sezione di Milano; [b] Dipartimento di Fisica, Università di Milano, Milano, Italy
[93] B.I. Stepanov Institute of Physics, National Academy of Sciences of Belarus, Minsk, Republic of Belarus
[94] National Scientific and Educational Centre for Particle and High Energy Physics, Minsk, Republic of Belarus
[95] Group of Particle Physics, University of Montreal, Montreal QC, Canada
[96] P.N. Lebedev Physical Institute of the Russian Academy of Sciences, Moscow, Russia
[97] Institute for Theoretical and Experimental Physics (ITEP), Moscow, Russia
[98] National Research Nuclear University MEPhI, Moscow, Russia
[99] D.V. Skobeltsyn Institute of Nuclear Physics, M.V. Lomonosov Moscow State University, Moscow, Russia
[100] Fakultät für Physik, Ludwig-Maximilians-Universität München, München, Germany
[101] Max-Planck-Institut für Physik (Werner-Heisenberg-Institut), München, Germany
[102] Nagasaki Institute of Applied Science, Nagasaki, Japan
[103] Graduate School of Science and Kobayashi-Maskawa Institute, Nagoya University, Nagoya, Japan
[104] [a] INFN Sezione di Napoli; [b] Dipartimento di Fisica, Università di Napoli, Napoli, Italy
[105] Department of Physics and Astronomy, University of New Mexico, Albuquerque NM, United States of America
[106] Institute for Mathematics, Astrophysics and Particle Physics, Radboud University Nijmegen/Nikhef, Nijmegen, Netherlands
[107] Nikhef National Institute for Subatomic Physics and University of Amsterdam, Amsterdam,

Netherlands  
[108] Department of Physics, Northern Illinois University, DeKalb IL, United States of America  
[109] Budker Institute of Nuclear Physics, SB RAS, Novosibirsk, Russia  
[110] Department of Physics, New York University, New York NY, United States of America  
[111] Ohio State University, Columbus OH, United States of America  
[112] Faculty of Science, Okayama University, Okayama, Japan  
[113] Homer L. Dodge Department of Physics and Astronomy, University of Oklahoma, Norman OK, United States of America  
[114] Department of Physics, Oklahoma State University, Stillwater OK, United States of America  
[115] Palacký University, RCPTM, Olomouc, Czech Republic  
[116] Center for High Energy Physics, University of Oregon, Eugene OR, United States of America  
[117] LAL, Univ. Paris-Sud, CNRS/IN2P3, Université Paris-Saclay, Orsay, France  
[118] Graduate School of Science, Osaka University, Osaka, Japan  
[119] Department of Physics, University of Oslo, Oslo, Norway  
[120] Department of Physics, Oxford University, Oxford, United Kingdom  
[121] $(a)$ INFN Sezione di Pavia; $(b)$ Dipartimento di Fisica, Università di Pavia, Pavia, Italy  
[122] Department of Physics, University of Pennsylvania, Philadelphia PA, United States of America  
[123] National Research Centre "Kurchatov Institute" B.P.Konstantinov Petersburg Nuclear Physics Institute, St. Petersburg, Russia  
[124] $(a)$ INFN Sezione di Pisa; $(b)$ Dipartimento di Fisica E. Fermi, Università di Pisa, Pisa, Italy  
[125] Department of Physics and Astronomy, University of Pittsburgh, Pittsburgh PA, United States of America  
[126] $(a)$ Laboratório de Instrumentação e Física Experimental de Partículas - LIP, Lisboa; $(b)$ Faculdade de Ciências, Universidade de Lisboa, Lisboa; $(c)$ Department of Physics, University of Coimbra, Coimbra; $(d)$ Centro de Física Nuclear da Universidade de Lisboa, Lisboa; $(e)$ Departamento de Fisica, Universidade do Minho, Braga; $(f)$ Departamento de Fisica Teorica y del Cosmos and CAFPE, Universidad de Granada, Granada (Spain); $(g)$ Dep Fisica and CEFITEC of Faculdade de Ciencias e Tecnologia, Universidade Nova de Lisboa, Caparica, Portugal  
[127] Institute of Physics, Academy of Sciences of the Czech Republic, Praha, Czech Republic  
[128] Czech Technical University in Prague, Praha, Czech Republic  
[129] Faculty of Mathematics and Physics, Charles University in Prague, Praha, Czech Republic  
[130] State Research Center Institute for High Energy Physics (Protvino), NRC KI, Russia  
[131] Particle Physics Department, Rutherford Appleton Laboratory, Didcot, United Kingdom  
[132] $(a)$ INFN Sezione di Roma; $(b)$ Dipartimento di Fisica, Sapienza Università di Roma, Roma, Italy  
[133] $(a)$ INFN Sezione di Roma Tor Vergata; $(b)$ Dipartimento di Fisica, Università di Roma Tor Vergata, Roma, Italy  
[134] $(a)$ INFN Sezione di Roma Tre; $(b)$ Dipartimento di Matematica e Fisica, Università Roma Tre, Roma, Italy  
[135] $(a)$ Faculté des Sciences Ain Chock, Réseau Universitaire de Physique des Hautes Energies - Université Hassan II, Casablanca; $(b)$ Centre National de l'Energie des Sciences Techniques Nucleaires, Rabat; $(c)$ Faculté des Sciences Semlalia, Université Cadi Ayyad, LPHEA-Marrakech; $(d)$ Faculté des Sciences, Université Mohamed Premier and LPTPM, Oujda; $(e)$ Faculté des sciences, Université Mohammed V, Rabat, Morocco  
[136] DSM/IRFU (Institut de Recherches sur les Lois Fondamentales de l'Univers), CEA Saclay (Commissariat à l'Energie Atomique et aux Energies Alternatives), Gif-sur-Yvette, France  
[137] Santa Cruz Institute for Particle Physics, University of California Santa Cruz, Santa Cruz CA, United States of America

[138] Department of Physics, University of Washington, Seattle WA, United States of America
[139] Department of Physics and Astronomy, University of Sheffield, Sheffield, United Kingdom
[140] Department of Physics, Shinshu University, Nagano, Japan
[141] Fachbereich Physik, Universität Siegen, Siegen, Germany
[142] Department of Physics, Simon Fraser University, Burnaby BC, Canada
[143] SLAC National Accelerator Laboratory, Stanford CA, United States of America
[144] [a] Faculty of Mathematics, Physics & Informatics, Comenius University, Bratislava; [b] Department of Subnuclear Physics, Institute of Experimental Physics of the Slovak Academy of Sciences, Kosice, Slovak Republic
[145] [a] Department of Physics, University of Cape Town, Cape Town; [b] Department of Physics, University of Johannesburg, Johannesburg; [c] School of Physics, University of the Witwatersrand, Johannesburg, South Africa
[146] [a] Department of Physics, Stockholm University; [b] The Oskar Klein Centre, Stockholm, Sweden
[147] Physics Department, Royal Institute of Technology, Stockholm, Sweden
[148] Departments of Physics & Astronomy and Chemistry, Stony Brook University, Stony Brook NY, United States of America
[149] Department of Physics and Astronomy, University of Sussex, Brighton, United Kingdom
[150] School of Physics, University of Sydney, Sydney, Australia
[151] Institute of Physics, Academia Sinica, Taipei, Taiwan
[152] Department of Physics, Technion: Israel Institute of Technology, Haifa, Israel
[153] Raymond and Beverly Sackler School of Physics and Astronomy, Tel Aviv University, Tel Aviv, Israel
[154] Department of Physics, Aristotle University of Thessaloniki, Thessaloniki, Greece
[155] International Center for Elementary Particle Physics and Department of Physics, The University of Tokyo, Tokyo, Japan
[156] Graduate School of Science and Technology, Tokyo Metropolitan University, Tokyo, Japan
[157] Department of Physics, Tokyo Institute of Technology, Tokyo, Japan
[158] Department of Physics, University of Toronto, Toronto ON, Canada
[159] [a] TRIUMF, Vancouver BC; [b] Department of Physics and Astronomy, York University, Toronto ON, Canada
[160] Faculty of Pure and Applied Sciences, and Center for Integrated Research in Fundamental Science and Engineering, University of Tsukuba, Tsukuba, Japan
[161] Department of Physics and Astronomy, Tufts University, Medford MA, United States of America
[162] Department of Physics and Astronomy, University of California Irvine, Irvine CA, United States of America
[163] [a] INFN Gruppo Collegato di Udine, Sezione di Trieste, Udine; [b] ICTP, Trieste; [c] Dipartimento di Chimica, Fisica e Ambiente, Università di Udine, Udine, Italy
[164] Department of Physics and Astronomy, University of Uppsala, Uppsala, Sweden
[165] Department of Physics, University of Illinois, Urbana IL, United States of America
[166] Instituto de Fisica Corpuscular (IFIC) and Departamento de Fisica Atomica, Molecular y Nuclear and Departamento de Ingeniería Electrónica and Instituto de Microelectrónica de Barcelona (IMB-CNM), University of Valencia and CSIC, Valencia, Spain
[167] Department of Physics, University of British Columbia, Vancouver BC, Canada
[168] Department of Physics and Astronomy, University of Victoria, Victoria BC, Canada
[169] Department of Physics, University of Warwick, Coventry, United Kingdom
[170] Waseda University, Tokyo, Japan
[171] Department of Particle Physics, The Weizmann Institute of Science, Rehovot, Israel

[172] Department of Physics, University of Wisconsin, Madison WI, United States of America
[173] Fakultät für Physik und Astronomie, Julius-Maximilians-Universität, Würzburg, Germany
[174] Fakultät für Mathematik und Naturwissenschaften, Fachgruppe Physik, Bergische Universität Wuppertal, Wuppertal, Germany
[175] Department of Physics, Yale University, New Haven CT, United States of America
[176] Yerevan Physics Institute, Yerevan, Armenia
[177] Centre de Calcul de l'Institut National de Physique Nucléaire et de Physique des Particules (IN2P3), Villeurbanne, France
[a] Also at Department of Physics, King's College London, London, United Kingdom
[b] Also at Institute of Physics, Azerbaijan Academy of Sciences, Baku, Azerbaijan
[c] Also at Novosibirsk State University, Novosibirsk, Russia
[d] Also at TRIUMF, Vancouver BC, Canada
[e] Also at Department of Physics & Astronomy, University of Louisville, Louisville, KY, United States of America
[f] Also at Department of Physics, California State University, Fresno CA, United States of America
[g] Also at Department of Physics, University of Fribourg, Fribourg, Switzerland
[h] Also at Departament de Fisica de la Universitat Autonoma de Barcelona, Barcelona, Spain
[i] Also at Departamento de Fisica e Astronomia, Faculdade de Ciencias, Universidade do Porto, Portugal
[j] Also at Tomsk State University, Tomsk, Russia
[k] Also at Universita di Napoli Parthenope, Napoli, Italy
[l] Also at Institute of Particle Physics (IPP), Canada
[m] Also at National Institute of Physics and Nuclear Engineering, Bucharest, Romania
[n] Also at Department of Physics, St. Petersburg State Polytechnical University, St. Petersburg, Russia
[o] Also at Department of Physics, The University of Michigan, Ann Arbor MI, United States of America
[p] Also at Centre for High Performance Computing, CSIR Campus, Rosebank, Cape Town, South Africa
[q] Also at Louisiana Tech University, Ruston LA, United States of America
[r] Also at Institucio Catalana de Recerca i Estudis Avancats, ICREA, Barcelona, Spain
[s] Also at Graduate School of Science, Osaka University, Osaka, Japan
[t] Also at Department of Physics, National Tsing Hua University, Taiwan
[u] Also at Institute for Mathematics, Astrophysics and Particle Physics, Radboud University Nijmegen/Nikhef, Nijmegen, Netherlands
[v] Also at Department of Physics, The University of Texas at Austin, Austin TX, United States of America
[w] Also at Institute of Theoretical Physics, Ilia State University, Tbilisi, Georgia
[x] Also at CERN, Geneva, Switzerland
[y] Also at Georgian Technical University (GTU),Tbilisi, Georgia
[z] Also at Ochadai Academic Production, Ochanomizu University, Tokyo, Japan
[aa] Also at Manhattan College, New York NY, United States of America
[ab] Also at Hellenic Open University, Patras, Greece
[ac] Also at Academia Sinica Grid Computing, Institute of Physics, Academia Sinica, Taipei, Taiwan
[ad] Also at School of Physics, Shandong University, Shandong, China
[ae] Also at Moscow Institute of Physics and Technology State University, Dolgoprudny, Russia
[af] Also at Section de Physique, Université de Genève, Geneva, Switzerland
[ag] Also at Eotvos Lorand University, Budapest, Hungary
[ah] Also at International School for Advanced Studies (SISSA), Trieste, Italy
[ai] Also at Department of Physics and Astronomy, University of South Carolina, Columbia SC, United States of America
[aj] Also at School of Physics and Engineering, Sun Yat-sen University, Guangzhou, China

[ak] Also at Institute for Nuclear Research and Nuclear Energy (INRNE) of the Bulgarian Academy of Sciences, Sofia, Bulgaria
[al] Also at Faculty of Physics, M.V.Lomonosov Moscow State University, Moscow, Russia
[am] Also at Institute of Physics, Academia Sinica, Taipei, Taiwan
[an] Also at National Research Nuclear University MEPhI, Moscow, Russia
[ao] Also at Department of Physics, Stanford University, Stanford CA, United States of America
[ap] Also at Institute for Particle and Nuclear Physics, Wigner Research Centre for Physics, Budapest, Hungary
[aq] Also at Flensburg University of Applied Sciences, Flensburg, Germany
[ar] Also at University of Malaya, Department of Physics, Kuala Lumpur, Malaysia
[as] Also at CPPM, Aix-Marseille Université and CNRS/IN2P3, Marseille, France
* Deceased


\else

\newpage
\begin{flushleft}
{\Large The ATLAS Collaboration}

\bigskip

M.~Aaboud$^\textrm{\scriptsize 135d}$,
G.~Aad$^\textrm{\scriptsize 86}$,
B.~Abbott$^\textrm{\scriptsize 113}$,
J.~Abdallah$^\textrm{\scriptsize 64}$,
O.~Abdinov$^\textrm{\scriptsize 12}$,
B.~Abeloos$^\textrm{\scriptsize 117}$,
R.~Aben$^\textrm{\scriptsize 107}$,
O.S.~AbouZeid$^\textrm{\scriptsize 137}$,
N.L.~Abraham$^\textrm{\scriptsize 149}$,
H.~Abramowicz$^\textrm{\scriptsize 153}$
% ...
\bigskip
\\
% ...
$^{12}$ Institute of Physics, Azerbaijan Academy of Sciences, Baku, Azerbaijan\\
% ...
$^{64}$ University of Iowa, Iowa City IA, United States of America\\
% ...
$^{86}$ CPPM, Aix-Marseille Universit{\'e} and CNRS/IN2P3, Marseille, France\\
% ...
$^{107}$ Nikhef National Institute for Subatomic Physics and University of Amsterdam, Amsterdam, Netherlands\\
% ...
$^{113}$ Homer L. Dodge Department of Physics and Astronomy, University of Oklahoma, Norman OK, United States of America\\
% ...
$^{117}$ LAL, Univ. Paris-Sud, CNRS/IN2P3, Universit{\'e} Paris-Saclay, Orsay, France\\
% ...
$^{135}$ $^{(a)}$ Facult{\'e} des Sciences Ain Chock, R{\'e}seau Universitaire de Physique des Hautes Energies - Universit{\'e} Hassan II, Casablanca; $^{(b)}$ Centre National de l'Energie des Sciences Techniques Nucleaires, Rabat; $^{(c)}$ Facult{\'e} des Sciences Semlalia, Universit{\'e} Cadi Ayyad, LPHEA-Marrakech; $^{(d)}$ Facult{\'e} des Sciences, Universit{\'e} Mohamed Premier and LPTPM, Oujda; $^{(e)}$ Facult{\'e} des sciences, Universit{\'e} Mohammed V, Rabat, Morocco\\
% ...
$^{137}$ Santa Cruz Institute for Particle Physics, University of California Santa Cruz, Santa Cruz CA, United States of America\\
% ...
$^{149}$ Department of Physics and Astronomy, University of Sussex, Brighton, United Kingdom\\
% ...
$^{153}$ Raymond and Beverly Sackler School of Physics and Astronomy, Tel Aviv University, Tel Aviv, Israel\\
% ...
$^{a}$ Also at Department of Physics, King's College London, London, United Kingdom\\
$^{b}$ Also at Institute of Physics, Azerbaijan Academy of Sciences, Baku, Azerbaijan\\
$^{c}$ Also at Novosibirsk State University, Novosibirsk, Russia\\
$^{d}$ Also at TRIUMF, Vancouver BC, Canada\\
$^{e}$ Also at Department of Physics {\&} Astronomy, University of Louisville, Louisville, KY, United States of America\\
% ...
\end{flushleft}

\fi

\end{document}